\documentclass[aps,prb,twocolumn,preprintnumbers,amsmath,amssymb,superscriptaddress]{revtex4-1}
\pdfoutput=1

\usepackage{graphicx}
\usepackage{bm}
\usepackage[usenames,dvipsnames]{xcolor}
\usepackage{hyperref}
\usepackage[tight]{subfigure}
\usepackage{verbatim}
\usepackage{units}
\usepackage{multirow}
\usepackage{enumitem}

\usepackage[english]{babel}
\renewcommand{\vec}[1]{\mathbf{#1}}

\newcommand{\vecG}[1]{\bm{#1}}

\newcommand{\bra}[1]{\langle #1|}
\newcommand{\ket}[1]{|#1 \rangle}

\newcommand{\Ham}{\mathcal{H}}

\usepackage[normalem]{ulem}

\usepackage{hyperref}
\hypersetup{colorlinks=true,linktoc=all,linkcolor=blue,citecolor=blue}

\begin{document}




\title{Spin-dependent thermoelectric effects in transport through a nanoscopic junction involving spin impurity}

\author{Maciej Misiorny}
 \email{misiorny@amu.edu.pl}
\affiliation{Peter Gr{\"u}nberg Institut PGI-2, Forschungszentrum J{\"u}lich, 52425 J{\"u}lich,  Germany}
\affiliation{JARA\,--\,Fundamentals of Future Information Technology, 52425 J{\"u}lich,  Germany}
\affiliation{Faculty of Physics, Adam Mickiewicz University, 61-614 Pozna\'{n}, Poland}

\author{J\'{o}zef Barna\'{s}}
\affiliation{Faculty of Physics, Adam Mickiewicz University, 61-614 Pozna\'{n}, Poland}
\affiliation{Institute of Molecular Physics, Polish Academy of Sciences, 60-179 Pozna\'{n}, Poland}

\date{\today}

\begin{abstract}
Conventional and spin-related  thermoelectric  effects in transport through a magnetic tunnel junction with
a large-spin impurity, such as a magnetic molecule or atom,
embedded into the corresponding barrier
are  studied theoretically in the linear-response regime. The impurity
is described by the giant spin Hamiltonian, with both uniaxial and transverse magnetic anisotropy taken into account.
Owing to the presence of transverse component of magnetic anisotropy, spin of a tunneling electron can be
reversed during scattering on the impurity, even in the low temperature regime. This reversal appears
due to exchange interaction of tunneling electrons with the magnetic impurity.
We calculate Seebeck and spin Seebeck coefficients, and analyze their dependence on various parameters of the spin impurity
and tunnel junction.
In addition, conventional and spin figures of merit, as well as the electronic contribution to heat conductance are considered.
We also show that pure spin current can be driven by a spin bias applied to the  junction with spin impurity, even if no electron transfer between the electrodes can take place. The underlying mechanism employs single-electrode tunneling processes (electrode-spin exchange interaction) and
the impurity as an intermediate reservoir of angular momentum.
\end{abstract}



\maketitle

\section{Introduction}

The potential success of novel spintronic nanoscopic devices relies on a
complete understanding of fundamental mechanisms governing the transport
of charge, spin and energy. In fact,  the thermoelectric and
thermomagnetic effects have been the subject of research in
condensed matter physics for nearly two
centuries.~\cite{Barnard_book} However, recently one observes a renaissance of interest
in the phenomena based on the interplay of transport of charge, spin and energy, especially in
nanoscopic systems.~\cite{VanHouten_Semicond.Sci.Technol.7/1992,Malen_Chem.Phys.Lett.491/2010,Dubi_Rev.Mod.Phys.83/2011,Goennenwein_NatureNanotechnol.7/2012,Bauer_NatureMater.11/2012,[{See
Chap.~9 (G.\,E.\,W. Bauer, `Spin caloritronics') in }]Maekawa_book_SC}

In general, the thermoelectric phenomena in  metals, such as the Seebeck
or Peltier effects (see Sec.~\ref{Sec:Background_thermoel}),  stem from
the electron-hole asymmetry.~\cite{Barnard_book,Bauer_NatureMater.11/2012} Whereas the
consequences of this fact were well understood for bulk and continuous
systems, the experiment of Smith \emph{et
al.}~\cite{Smith_Phys.Rev.B22/1980} demonstrating the thermoelectric
effect in a tunnel junction (formed by two metallic electrodes separated
by an oxide barrier)  proved that this is valid in principle also for
mesoscopic systems consisting of  discrete subsystems -- each of them
being in local equilibrium, but not necessarily in equilibrium with other subsystems. The
idea was further developed by Johnson and
Silsbee,~\cite{Johnson_Phys.Rev.B35/1987} who suggested that if at least
one of the electrodes is ferromagnetic and
the junction is out of thermodynamic equilibrium, not only do thermally stimulated
voltages and heat transport arise, but also electrically
and thermally induced magnetization currents can appear. Furthermore,
they also predicted the reciprocal effects, i.e. `magnetically' stimulated electrical and thermal currents.
Independently, a theory of linear electrical and thermal transport
between two metallic reservoirs  interconnected \emph{via}  ideal leads
to an arbitrary disordered system was developed by Sivan and
Imry.~\cite{Sivan_Phys.Rev.B33/1986}

The interest in thermoelectric properties of nanoscopic
systems, however, has been only awakened by first experiments involving
quantum
dots,~\cite{Staring_Europhys.Lett.22/1993,Molenkamp_Semicond.Sci.Technol.9/1994,Godijn_Phys.Rev.Lett.82/1999,Scheibner_Phys.Rev.Lett.95/2005,Scheibner_Phys.Rev.B75/2007}
which have been followed by numerous theoretical works covering  the
limits of both
weak~\cite{Beenakker_Phys.Rev.B46/1992,Turek_Phys.Rev.B65/2002,Koch_Phys.Rev.B70/2004,Dubi_Phys.Rev.B79/2009,Leijnse_Phys.Rev.B82/2010,Wierzbicki_Phys.Lett.A375/2011,Sanchez_Phys.Rev.B84/2011,Muralidharan_Phys.Rev.B85/2012,Sanchez_Phys.Rev.Lett.110/2013,Svensson_NewJ.Phys.15/2013}
and
strong~\cite{Boese_Europhys.Lett.56/2001,Andreev_Phys.Rev.Lett.86/2001,Matveev_Phys.Rev.B66/2002,Krawiec_Phys.Rev.B73/2006,Costi_Phys.Rev.B81/2010,Nguyen_Phys.Rev.B82/2010,Weymann_Phys.Rev.B88/2013,Zitko_NewJ.Physics15/2013}
tunnel coupling between the dot and electrodes. An important practical
aspect accompanying research on thermoelectric effects is their potential
significance for harnessing power dissipated as heat, and thus reducing
the loss of energy.~\cite{Bell_Science321/2008} In this respect, a
prospective candidate as heat-voltage converters seem to be single
molecules. Experiments on molecular junctions employing a scanning
tunneling microscope (STM) setup and  comprising up to a few
molecules~\cite{Reddy_Science315/2007,Baheti_NanoLett.8/2008,Malen_NanoLett.9/2009_1164,Malen_NanoLett.9/2009_3406,Tan_App.Phys.Lett.96/2010,Widawsky_NanoLett.12/2011,Yee_NanoLett.11/2011,Evangeli_NanoLett.13/2013}
have shown that such systems remain  thermoelectrically responsive at
temperatures as high as room temperature, and their specific
thermoelectric properties can be to some extent tailored by chemical
engineering. This, in turn,  has also triggered a significant interest
in theoretical description of thermoelectric transport through molecular
junctions.~\cite{Paulsson_Phys.Rev.B67/2003,Segal_Phys.Rev.B72/2005,Bergfield_NanoLett.9/2009,Bergfield_Phys.Rev.B79/2009,Dubi_NewJ.Phys.15/2013}
It has been suggested that the efficiency of such devices can be
improved due to a violation of the Wiedemann-Franz law
occurring as a consequence of the system's energy quantization and
Coulomb
interactions.~\cite{Boese_Europhys.Lett.56/2001,Krawiec_Phys.Rev.B73/2006,Vavilov_Phys.Rev.B72/2005,Ahmadian_Phys.Rev.B72/2005,Murphy_Phys.Rev.B78/2008,Kubala_Phys.Rev.Lett.100/2008}

Since electrons -- apart from charge -- possess also a spin degree of freedom,
one should thus expect the interplay between spin and heat currents
leading to some interesting thermoelectric
phenomena.~\cite{Goennenwein_NatureNanotechnol.7/2012}  In general, these
can be further distinguished into independent electron and collective
effects.~\cite{Bauer_NatureMater.11/2012} The former group comprises
effects which can be explained by a model of two independent spin-channels,
and thus it is  limited to systems where the
spin-flip diffusion length of conduction electrons is sufficiently long
with respect to the system's length scale. On the other hand, in the case of
the effects belonging to the second
group, the  spin
currents are not simply only due to a particle current but they are also
carried by magnon
excitations.~\cite{Xiao_Phys.Rev.B81/2010,Adachi_Phys.Rev.B83/2011,Adachi_Rep.Prog.Phys.76/2013}
Consequently, unlike the independent electron effects which are limited
to metallic systems, the collective thermoelectric effects can arise not
only in metallic
ferromagnets,~\cite{Uchida_Nature455/2008,Uchida_SolidStateCommun.150/2010,Bosu_Phys.Rev.B83/2011}
but also in semiconducting
ferromagnets~\cite{Jaworski_NatureMater.9/2010} or even in insulating
magnets.~\cite{Uchida_NatureMater.9/2010,Uchida_Appl.Phys.Lett.97/2010}

The spin-dependent thermoelectric effects, which are the subject of the present
paper,  have been experimentally studied in a variety of nanoscopic
systems, such as  magnetic tunnel
junctions,~\cite{Bauer_NatureMater.11/2012,LeBreton_Nature475/2011,Walter_NatureMater.10/2011,Lin_NatureCommun.3/2012}
nanopillars,~\cite{Liebing_Phys.Rev.Lett.107/2011,Dejene_Phys.Rev.B86/2012,Flipse_NatureNanotechnol.7/2012}
nonlocal spin-valve
devices,~\cite{Bakker_Phys.Rev.Lett.105/2010,Erekhinsky_Appl.Phys.Lett.100/2012}
as well as
multilayered~\cite{Gravier_Phys.Rev.B73/2006,Gravier_Phys.Rev.B73BR/2006} and
granular~\cite{Shi_Phys.Rev.B54/1996} systems. Moreover, such effects
have also been extensively studied theoretically in magnetic tunnel
junctions,~\cite{Wang_Phys.Rev.B63/2001,McCann_Phys.Rev.B66/2002,McCann_Appl.Phys.Lett.81/2002,Jansen_Phys.Rev.B85/2012}
local~\cite{Hatami_Phys.Rev.B79/2009} and
nonlocal~\cite{Slachter_NaturePhys.6/2010} spin valves, quantum
dots,~\cite{Swirkowicz_Phys.Rev.B80/2009,Trocha_Phys.Rev.B85/2012}
wires,~\cite{Rejec_Phys.Rev.B65/2002}
wells,~\cite{Rothe_Phys.Rev.B86/2012} or even in single-molecule-magnet
junctions.~\cite{Wang_Phys.Rev.Lett.105/2010,Zhang_Appl.Phys.Lett.97/2010}
Interestingly enough, it has been predicted, for example,  that
spin-polarized thermoelectric heat currents can reverse the
magnetization direction of a
ferromagnet,~\cite{Hatami_Phys.Rev.Lett.99/2007} which appears due to the
spin-transfer torque associated with purely thermal currents. This effect has
been later confirmed by experiment.~\cite{Yu_Phys.Rev.Lett.104/2010}
Recently, a thermoelectric equivalent of spin accumulation, i.e. spin
heat accumulation, manifested as different effective temperatures
for the spin-up and spin-down electrons, has been observed in a
nanopillar spin
valve.~\cite{Dejene_NaturePhys.9/2013,Vera-Marun_arXiv:1308.3365/2013}

In the present paper we focus on spin-dependent  thermoelectric effects
that can arise in \emph{linear-response} transport through a nanoscopic
junction in which an impurity of spin $S>1/2$ is embedded into a barrier. Unlike in the
case considered by Johnson and Silsbee,~\cite{Johnson_Phys.Rev.B35/1987}
spins of conduction electrons tunneling through the junction
can be reversed owing to scattering on the impurity. Such spin-flip
scattering processes lead to exchange of angular momentum between the
conduction electrons and the impurity, which in turn allows for the
control of spatial orientation of the impurity's
spin.~\cite{Misiorny_Phys.Rev.B75/2007,Fransson_NanoLett.9/2009,Delgado_Phys.Rev.Lett.104/2010}

In the linear response regime and when no excitations are permitted, the
spin exchange processes can  result in transitions of the impurity only
between degenerate spin states whose angular momentum differs by the quantum of angular momentum
$\hbar$. For a \emph{spin-isotropic} impurity, where all $2S+1$ spin states are degenerate, this means that all  these
states can in principle contribute to transport. However, usually a
spatial symmetry of a high-spin system is broken by the presence of
environment, e.g., as for a magnetic atom placed on a
surface,~\cite{Brune_Surf.Sci.603/2009} which renders the system
\emph{spin-anisotropic}. If only the  \emph{uniaxial}  anisotropy exists, the two
ground spin states are separated by an energy barrier. At
sufficiently low temperatures,
i.e. lower than the \emph{zero-field splitting} energy between the ground and first excited doublets -- being also the largest excitation energy between two consecutive states,
 the impurity occupies then only the ground
state spin doublet, and no direct transitions are allowed between the doublet
ground states. In consequence, in linear response regime only
spin-conserving transport processes are possible. The situation changes when
the \emph{transverse} magnetic anisotropy is present in the system, as it
allows for mixing of states with different $S_z$ numbers. In particular,
for a half-integer spin one obtains a ground state Kramers' doublet, as follows from time-inversion symmetry.

In this paper we consider the situation when no energy excitations of the impurity are admitted. Thus, when the impurity is \emph{anisotropic}, the temperature is
limited to the thermal energies smaller than the impurity excitation energy, and in particular the zero-field splitting energy.
 Accordingly, only ground state doublet is involved in the
linear-response regime.
On the other hand,
when the impurity is \emph{isotropic}, then all impurity states are degenerate and all are involved in transport in the linear-response regime. Thus, the above temperature restriction
becomes irrelevant, similarly as in the case of a junction with no impurity.
In Sec.~\ref{Sec:Background_thermoel} we provide some background on thermoelectric phenomena. Description of thermoelectricity in transport through a junction
with spin-impurity in the barrier is presented in Sec.~\ref{Sec:Transport_theory}. Numerical results and their discussion are given in Sec.~\ref{Sec:Results},
which is  followed by the section comprising final conclusions (Sec.~\ref{Sec:Conclusions}).

\section{\label{Sec:Background_thermoel}Background on thermoelectric phenomena}

Before introducing the model system to be considered in this paper and calculating the
thermoelectric parameters of interest,
we find it instructive to present some fundamental concepts regarding
the conventional thermoelectricity (for a more detailed discussion see,
e.g.,  Refs.~[\onlinecite{Barnard_book,Mahan_book,diVentra_book,Blundell_book_thermal,Dubi_Rev.Mod.Phys.83/2011}]),
and then their generalization to the corresponding spin thermoelectric phenomena. For this purpose, let's consider a  tunnel junction
in which two metallic electrodes (reservoirs of electrons) are separated by a tunnel barrier.
First, we consider the case when spin voltage is irrelevant (conventional thermoelectricity), and then we also include
the spin voltage (spin thermoelectricity).

\subsection{Conventional thermoelectricity}

When a constant \emph{voltage} $\delta V$ and \emph{thermal} $\delta T$ bias is maintained across the
junction, it results in a stationary net flow of charge and
heat.~\cite{Smith_Phys.Rev.B22/1980,Johnson_Phys.Rev.B35/1987,Houten_Semicond.Sci.Technol.7/1992,Houten_Phys.Today49/1996}
Moreover, since charge and energy are in fact both carried by electrons (we do not consider here energy carried by phonons),
the corresponding \emph{charge} $I_\text{C}$ and \emph{heat} $I_\text{Q}$ currents are  related, leading to a variety of
thermoelectric effects and relations.

In order to reveal relation between charge and heat transport, let's first focus on the transport of electrons under
isothermal conditions, $\delta T =0$. In such a case, the charge current $I_\text{C}$ is driven exclusively by a
voltage bias $\delta V$, and the relevant transport coefficient is the well-known \emph{electrical conductance} $G$,
    \begin{equation}\label{Eq:G_def}
    G=\left(\frac{I_\text{C}}{\delta V}\right)_{\!\!\delta T=0}
    .
    \end{equation}
However, even though $\delta T=0$, there is a  heat current associated with the electrical current, and this
phenomenon is referred to as \emph{the Peltier effect}.
The relevant relation between  heat $I_\text{Q}$ and charge $I_\text{C}$ currents is then described by
the \emph{Peltier coefficient} $\Pi$,

    \begin{equation}\label{Eq:Pi_def}
    \Pi=\left(\frac{I_\textrm{Q}}{I_\text{C}}\right)_{\!\!\delta T=0}
    .
    \end{equation}

Another limiting situation appears when the heat transfer through the system occurs due to thermal bias
in the absence of a charge current, $I_\text{C}=0$.
The latter condition can be easily achieved when the system is in an electrically open circuit.
The electronic contribution to the \emph{thermal conductance}~$\kappa$ is then defined as
    \begin{equation}\label{Eq:kappa_def}
    \kappa=\left(\frac{I_\textrm{Q}}{\delta T}\right)_{\!\!I_\text{C}=0}
    .
    \end{equation}
Although the resultant flow of electrons is now equal to zero, a voltage difference between the two reservoirs
appears as a result of thermal gradient. This phenomenon is known as the \emph{Seebeck effect} and is
characterized by the \emph{thermopower} (\emph{Seebeck coefficient}) $\mathcal{S}$,
    \begin{equation}\label{Eq:S_def}
    \mathcal{S}=-\left(\frac{\delta V}{\delta T}\right)_{\!\!I_\text{C}=0}
    .
    \end{equation}
Note that in order  to achieve the condition $I_C=0$ in an electrically closed system, one needs to apply
an external voltage compensating the charge current due to the temperature gradient.

Finally, the overall thermoelectric efficiency of  a system is described by the so-called \emph{figure of merit} \text{ZT},
    \begin{equation}\label{Eq:ZT_def}
    \textrm{ZT}=\frac{\mathcal{S}^2GT}{\kappa}
    ,
    \end{equation}
which is a dimensionless quantity expressed in terms of the experimentally measurable coefficients $G$, $\kappa$ and $S$.
Note that $\kappa$
in Eq.(\ref{Eq:ZT_def}) generally includes also the thermal conductance due to phonons, which is not considered here.

\subsection{Spin thermoelectricity}

The concepts briefly described above can be further generalized to the transport model based on two nonequivalent
spin channels.~\cite{Bauer_NatureMater.11/2012} Let's note first that charge transport in ferromagnetic conductors
is generally associated with a spin current. Second, the electrochemical potentials for
spin-up  and spin-down electrons can be different in the vicinity of an interface between ferromagnetic and nonmagnetic materials (i.e. up to distances
of the order of the spin-flip diffusion length)  when  the rate of electron scattering without spin flip is significantly
larger than the spin-flip rate. This appears as spin accumulation at the interface.~\cite{Son_Phys.Rev.Lett.58/1987,Valet_Phys.Rev.B48/1993}
The spin accumulation (spin-dependent
electrochemical potentials), in turn, led to the concept of the
so-called \emph{spin bias} $\delta V_\text{S}$. Accordingly, the difference in electrochemical potentials of the two
electrodes in the spin-$\sigma$ channel, $\delta V_\sigma$,  can be written
as $\delta V_\sigma=\delta V+\eta_\sigma\delta V_\text{S}$, with $\eta_{\uparrow(\downarrow)}=\pm1$. Thus, with the
use of electrical and spin bias one can independently control electric and spin currents. Moreover, in certain
situations one can drive  pure spin current, i.e. spin current which is not associated with any charge current.
The occurrence of \emph{spin} currents $I_\text{S}$ initiated  the concept of \emph{spin} counterparts of the thermoelectric effects discussed above.

Under isothermal conditions  both charge $I_\text{C}$ and spin $I_\text{S}$ currents can be controlled independently by voltage $\delta V$ and spin
bias  $\delta V_\text{S}$.
Hence, one can define a generalized conductance matrix $\bm{G}$ as~\cite{Swirkowicz_Phys.Rev.B80/2009}
    \begin{equation}\label{Eq:G_S_def0}
    \renewcommand{\arraystretch}{2}
    \bm{G}
    \equiv
        \begin{pmatrix}
        G  & G^\text{m}
        \\[4pt]
       G_\text{S}^\text{m} & G_\text{S}^{}
        \end{pmatrix}
        =
        \begin{pmatrix}
        \left(
        \dfrac{I_\text{C}}{\delta V}
        \right)_{\!\!\begin{subarray}{l}\delta T=0\\\delta V_S=0\end{subarray}}
        &
        \left(
         \dfrac{I_\text{C}}{\delta V_\text{S}}
        \right)_{\!\!\begin{subarray}{l}\delta T=0\\\delta V=0\end{subarray}}
                    \\[6pt]
       \left(
        \dfrac{I_\text{S}}{\delta V}
        \right)_{\!\!\begin{subarray}{l}\delta T=0\\\delta V_S=0\end{subarray}}
        &
        \left(
        \dfrac{I_\text{S}}{\delta V_\text{S}}
        \right)_{\!\!\begin{subarray}{l}\delta T=0\\\delta V=0\end{subarray}}
        \end{pmatrix}
        .
   \end{equation}
Furthermore, in addition to the conventional Peltier coefficient $\Pi$, one can introduce a \emph{spin Peltier coefficient} $\Pi_\text{S}$,
    \begin{equation}\label{Eq:Pi_S_def}
    \Pi=\left(\frac{I_\text{Q}}{I_\text{C}}\right)_{\!\!\begin{subarray}{l}\delta T=0\\\delta V_\text{S}=0\end{subarray}}
    \quad
    \text{and}
    \quad
    \Pi_\text{S}=\left(\frac{I_\text{Q}}{I_\text{S}}\right)_{\!\!\begin{subarray}{l}\delta T=0\\\delta V=0\end{subarray}}
    .
    \end{equation}
The former coefficient describes a heat flow associated with an electrical current in the absence of spin voltage,
whereas the latter one represents the heat current associated with a  spin current for a zero voltage bias. Note,
the definition of the Peltier coefficient $\Pi$ is equivalent to that given by Eq.~(\ref{Eq:Pi_def})
providing spin accumulation is disregarded and there is no spin voltage.

The definition of thermal conductivity in Eq.~(\ref{Eq:kappa_def}) holds also in the present situation, when the
 additional constraint $\delta V_\text{S}=0$ is imposed,
    \begin{equation}\label{Eq:kappa_S}
    \kappa=\left(\frac{I_\text{Q}}{\delta T}\right)_{\!\!\begin{subarray}{l}I_\text{C}=0\\\delta V_\text{S}=0\end{subarray}}
    .
    \end{equation}
Interestingly, if spin accumulation can arise in the system, thermal bias can induce not only an electrical voltage $\delta V$, but also
a spin voltage $\delta V_S$. The latter effect is referred to as the \emph{spin Seebeck effect}. Consequently, along with the
conventional thermopower $\mathcal{S}$, one can formally define the \emph{spin thermopower} $\mathcal{S}_\text{S}$,~\cite{Dubi_Phys.Rev.B79/2009}
    \begin{equation}\label{Eq:S_S_def}
    \mathcal{S}=-\left(\frac{\delta V}{\delta T}\right)_{\!\!\begin{subarray}{l}I_\text{C}=0\\\delta V_\text{S}=0\end{subarray}}
    \quad
    \text{and}
    \quad
    \mathcal{S}_\text{S}=-\left(\frac{\delta V_\text{S}}{\delta T}\right)_{\!\!\begin{subarray}{l}I_\text{S}=0\\\delta V=0\end{subarray}}
    .
    \end{equation}
To complete the discussion of spin-dependent effects, we note that a spin analog of the figure of merit
$\text{ZT}_\text{S}$, see  Eq.~(\ref{Eq:ZT_def}), can be used to characterize the spin  thermoelectric efficiency of a system,
    \begin{equation}\label{Eq:ZT_S_def}
    \textrm{ZT}_\text{S}
    =
    \frac{2|e|}{\hbar}
    \frac{\mathcal{S}_\text{S}^2|G_\text{S}^{\mbox{}}|T}{\kappa}
    ,
    \end{equation}
with $G_\text{S}$ defined in Eq.~(\ref{Eq:G_S_def0}) and the thermal conductivity~$\kappa$ given by Eq.~(\ref{Eq:kappa_S}).

When a system is in an open circuit and there are no spin relaxation processes, then neither charge  nor spin
current can  flow through the system. Thus, one may alternatively define the thermoelectric coefficients for
$I_\text{C}=0$ and $I_\text{S}=0$,\cite{Swirkowicz_Phys.Rev.B80/2009} i.e.
    \begin{equation}
    \mathcal{S}=-\left(\frac{\delta V}{\delta T}\right)_{\!\!\begin{subarray}{l}I_\text{C}=0\\I_\text{S}=0\end{subarray}}
    \quad
    \text{and}
    \quad
    \mathcal{S}_\text{S}=-\left(\frac{\delta V_\text{S}}{\delta T}\right)_{\!\!\begin{subarray}{l}I_\text{C}=0\\I_\text{S}=0\end{subarray}}
    \end{equation}
for the charge and spin thermopowers, and also
    \begin{equation}
    \kappa=\left(\frac{I_\text{Q}}{\delta T}\right)_{\!\!\begin{subarray}{l}I_\text{C}=0\\I_\text{S}=0\end{subarray}}
    .
    \end{equation}
for the heat conductance.
Since to have zero spin current one needs to apply a spin voltage, the charge thermopower may be different
from that determined from the conventional formula. Note that this difference appears only when spin relaxation
is slow. In the following we will use the definitions (\ref{Eq:S_S_def}).

\section{\label{Sec:Transport_theory}Thermoelectricity in a magnetic junction with spin impurity}

\subsection{Theoretical model}

\begin{figure*}[t]
   \includegraphics[scale=0.55]{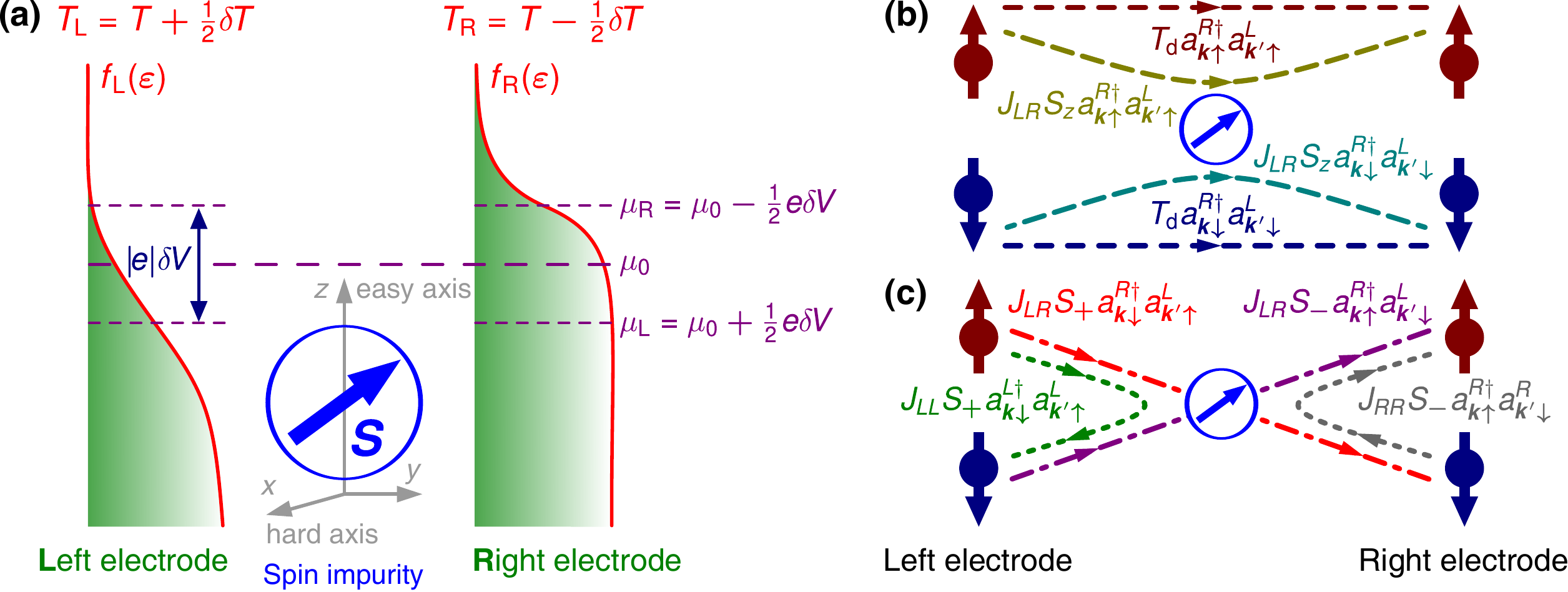}
    \caption{
    (color online)
    (a) Schematic depiction of the system under consideration. Transport of electrons between the left and right
    electrodes ($q=L,R$) appears due to an external bias voltage $\delta V$ ($\mu_0$ is the  electrochemical potential
    at equilibrium), and/or due to the difference $\delta T$ between electrodes' temperatures $T_q$. Note that for the
    sake of clarity we assume here that spin accumulation is absent, and consequently $\delta V_\text{S}=0$.  Different
    temperatures of electrodes are delineated here  with the use of the Fermi-Dirac distribution functions
    $f_q(\varepsilon)$, which are smeared out dissimilarly around the electrochemical potentials $\mu_q$ of the electrodes.
    Right panel represents graphically some examples of different possible \emph{spin-conserving} (b) and \emph{spin-flip}
    (c) electron tunneling processes. In particular, we specify that fine-dashed lines in (b) stand for direct tunneling of
    electrons between the electrodes, while dotted lines in (c) symbolize single-electrode  tunneling processes.
        }
    \label{Fig:1}
\end{figure*}

The system to be considered in the following consists of two metallic -- generally ferromagnetic -- electrodes separated by an insulating barrier. Tunneling
of electrons between the electrodes can appear  either due to applied voltage (electric or spin one) or due to a thermal
bias, see Fig.~\ref{Fig:1}(a). Experimentally, the system can be either a simple planar  magnetic tunnel junction or
a setup involving an STM tip as one of the electrodes. Additionally, we assume that  a magnetic impurity is embedded in
the barrier between the
electrodes,~\cite{Gregory_Phys.Rev.Lett.68/1992,Ralph_Phys.Rev.Lett.72/1994,Park_Nature417/2002,Liang_Nature417/2002,Hirjibehedin_Science317/2007,Meier_Science320/2008,Kahle_NanoLett.12/2012} which scatters electrons traversing the barrier. The total Hamiltonian of the system $\Ham$ thus consists of three terms, $\Ham=\Ham_\text{imp}+\Ham_\text{el}+\Ham_\text{T}$, representing the impurity, electrodes, and electron tunneling processes, respectively.

Here, we focus mainly on magnetic impurities characterized by a large spin number $S$, $S>1/2$, whose behavior is
dominated by the presence of magnetic anisotropy. In general, basic features of a large-spin magnetic impurity,
represented by a spin operator $\vec{S}=(S_x,S_y,S_z)$, are captured by the \emph{giant-spin} Hamiltonian,~\cite{Gatteschi_book}
    \begin{equation}\label{Eq:Ham_S}
    \Ham_\text{imp}
    =
    -DS_z^2 + \frac{E}{2}\big(S_+^2+S_-^2\big)
    ,
    \end{equation}
where the first and second terms denote the \emph{uniaxial} and \emph{transverse} magnetic anisotropy, respectively,
with $D$ and $E$ standing for the corresponding anisotropy constants and $S_\pm=S_x\pm iS_y$. Since  we are interested
here in  magnetic impurities capable of information storage,  we assume an energy barrier
for spin reversal, i.e. $D>0$. In addition, also the transverse anisotropy constant can be assumed to be
positive, $E>0$, and the two magnetic anisotropy constants satisfy the condition\cite{Gatteschi_book} $0\leqslant E/D\leqslant1/3$.
Next, we assume  a \emph{half-integer} spin $S$. Due to the presence of transverse magnetic
anisotropy, each of the $2S+1$ eigenstates $\ket{\chi_m}$  of the impurity  Hamiltonian~(\ref{Eq:Ham_S}), $\Ham_\text{imp}\ket{\chi_m}=E_{\chi_m}\ket{\chi_m}$,
is then a linear combination of the eigenstates $\ket{m}$ of the spin operator $S_z$. Note the notation we use for the eigenstates $\ket{\chi_m}$,
with the subscript $m$ corresponding to the $S_z$ component of highest weight in the state $\ket{\chi_m}$, i.e. $\lim_{E\rightarrow0}\ket{\chi_m}=\ket{m}$.
Moreover, the eigenstates $\ket{\chi_m}$ are twofold degenerate (Kramers' doublets) and form two uncoupled
sets\cite{Romeike_Phys.Rev.Lett.96/2006,Romeike_Phys.Rev.Lett.96/2006_TranspSpetr} $\big\{\ket{\chi_{\pm S\mp2k}}\big\}_{k=0,1,\ldots,S-1/2}$.
Thus, any eigenstate $\ket{\chi_m}$, for $m=-S,\dots,S$, can generally be written as
    \begin{equation}
    \ket{\chi_m}
    =
    \sum_{k\in\mathbb{Z},\, |m+2k|\leqslant S}
    \langle m+2k|\chi_m\rangle
    \ket{m+2k},
    \end{equation}
 where $\langle m+2k|\chi_m\rangle$ represents the overlap of the state \mbox{$\ket{m+2k}$} with the eigenstate $\ket{\chi_m}$.
 In particular, the eigenstates constituting the ground state Kramers' doublet take the form
    \begin{equation}
    \left\{
        \begin{aligned}
        &
        \ket{\chi_{-S}}
        =
        \sum_{k=0}^{S-1/2}
        \langle -S+2k|\chi_m\rangle
        \ket{-S+2k},
                \\
        &
        \ket{\chi_{S}}
        =
        \sum_{k=0}^{S-1/2}
        \langle S-2k|\chi_m\rangle
        \ket{S-2k},
        \end{aligned}
    \right.
    \end{equation}
 from which one concludes that  the system's spin can be trapped in one of two distinguishable spatial configurations
 with respect to the $z$th axis, referred to also as the system's \emph{easy axis}.\footnote{Note that for an \emph{integer}
 spin $S$ the ground state would be split even in the absence of an external magnetic field, and as  a symmetric admixture of
 states $\protect\ket{\pm S},\protect\ket{\pm S\pm 2},\ldots,\protect\ket{0}$ it would prefer orientation in the plane
 perpendicular to the system's easy axis.} In the following, we will use the index $m$ only when necessary to avoid any confusion.

Electrodes are treated as reservoirs of itinerant and noninteracting electrons,  and are described by the Hamiltonian
    \begin{equation}\label{Eq:Ham_el}
     \Ham_\textrm{el}=
     \sum_{q\vec{k}\sigma}
     \varepsilon_{q\vec{k}\sigma} a_{\vec{k}\sigma}^{q\dag} a_{\vec{k}\sigma}^q,
     \end{equation}
with  $\varepsilon_{q\vec{k}\sigma}$ denoting  the conduction electron energy in the $q$th electrode ($q=L$ for the left and $q=R$
for the right electrode, respectively),
$\vec{k}$ standing for a wave vector,  and $\sigma$ being the electron spin index. Furthermore,
$a_{\vec{k}\sigma}^{q\dag}$ ($a_{\vec{k}\sigma}^{q}$) is the relevant electron creation (annihilation) operator for
the $q$th electrode. Generally, both electrodes are  characterized by a \emph{spin-dependent} density of states
(DOS), $\sum_\vec{k}\delta(\varepsilon_{\vec{k}\sigma}^q-\omega)=\rho_\sigma^q(\omega)$. Importantly, for the problem
under discussion DOS of at least one of the electrodes has to be  energy-dependent and asymmetric around the Fermi level in order to obtain a nonzero thermopower.
 The magnetic properties of the $q$th electrode will be described by the corresponding  \emph{spin polarization coefficient}, $P_q$,
 defined at the system's Fermi level $\mu_0$ as
    \begin{equation}
    P_q
    =
    \frac{\rho_\uparrow^q(\mu_0)-\rho_\downarrow^q(\mu_0)}{\rho_\uparrow^q(\mu_0)+\rho_\downarrow^q(\mu_0)}
    .
    \end{equation}

Finally, electron tunneling processes between the electrodes are described by the Appelbaum
Hamiltonian,\cite{Appelbaum_Phys.Rev.Lett.17/1966,Appelbaum_Phys.Rev.154/1967,Kim_Phys.Rev.Lett.92/2004,Nussinov_Phys.Rev.B68/2003,Misiorny_Phys.Rev.B75/2007,Fransson_NanoLett.9/2009}
    \begin{equation}\label{Eq:Ham_T}
    \Ham_\textrm{T}
    =\!
    \sum_{q\vec{k}\vec{k}'\alpha}
    \!\!
    \Big\{
    T_\textrm{d}\,
    a_{\vec{k}\alpha}^{q\dag}
    a_{\vec{k}'\alpha}^{\bar{q}}
    +
    \sum_{q^\prime\beta}
    \!
    J_{qq^\prime}\,
    \vecG{\sigma}_{\alpha\beta}\cdot\vec{S}\:
    a_{\vec{k}\alpha}^{q\dag}
    a_{\vec{k}'\beta}^{q^\prime}
    \Big\}.
    \end{equation}
In the equation above $\overline{q}$ should be understood as $\overline{L}\equiv R$ and
$\overline{R}\equiv L$. Furthermore, $\vecG{\sigma}=(\sigma_x,\sigma_y,\sigma_z)$ and $\sigma_i$ ($i=x,y,z$) denote the Pauli matrices.
The first term of Eq.~(\ref{Eq:Ham_T}) represents direct tunneling of electrons between the electrodes, while the
second term  takes into account the fact that during tunneling  an electron can interact magnetically with the impurity
 either \emph{via} exchange coupling or  direct dipolar interactions,~\cite{Hirjibehedin_Science317/2007}  see Fig.~\ref{Fig:1}(b)-(c).
The former processes are then described by the tunneling parameter $T_\text{d}$, whereas the latter ones by the exchange
parameter $J_{qq^\prime}$, with both the parameters assumed to be real, isotropic, and independent of  energy and electrodes' spin polarization.
It is convenient to introduce the following parameterization for $T_\text{d}$ and $J_{qq^\prime}$:
$T_\text{d}=\alpha_\text{d}K$ and  $J_{qq^\prime}=\nu_q\nu_{q^\prime}J$,
with $\nu_q$ being a dimensionless factor quantifying the coupling between the impurity's spin and the $q$th electrode, and $J=\alpha_\text{ex}K$.
Thus, $K$ becomes the key, experimentally relevant parameter,\cite{Nussinov_Phys.Rev.B68/2003,Fransson_NanoLett.9/2009}
whereas $\alpha_\text{ex}/\alpha_\text{d}$ establishes a relationship between  the processes of direct electron tunneling
and those during which the spin of a tunneling electron can be reversed.

\subsection{\label{Sec:Transport_charcter}Transport characteristics}

I the following we assume weak coupling between the electrodes and the impurity. Charge, spin and energy transport can be then
 described within the approach based on the
corresponding master equation. Balance of respective  flows associated with tunneling of electrons ($e<0$)
out/to each electrode gives the relevant currents in the following form (see also Appendix A for
a more explicit
form of these expressions):
\begin{itemize}[leftmargin=*]
\item
\emph{charge current} $I_\text{C}=\big(I_\text{C}^L-I_\text{C}^R\big)/2$
    \begin{align}\label{Eq:I_C_def}
    I_\text{C}^q
    =
    e
    \sum_{\vec{k}\vec{k}^\prime}
    \sum_{\sigma\sigma^\prime}
    \sum_{\chi\chi^\prime}
    \mathcal{P}_\chi
    \Big\{
    &
    \mathcal{I}^{\ket{q\vec{k}\sigma,\chi}}_{\ket{\overline{q}\vec{k}^\prime\!\sigma^\prime,\chi^\prime}}
    -
    \mathcal{I}^{\ket{\overline{q}\vec{k}\sigma,\chi}}_{\ket{q\vec{k}^\prime\!\sigma^\prime,\chi^\prime}}
    \Big\}
    ,
    \end{align}
\item
 \emph{spin current} $I_\text{S}=\big(I_\text{S}^L-I_\text{S}^R\big)/2$
    \begin{align}\label{Eq:I_S_def}
    \hspace*{-2pt}
    I_\text{S}^q
    =\ &
    \frac{\hbar}{2}
    \sum_{q^\prime}
    \sum_{\vec{k}\vec{k}^\prime}
    \sum_{\alpha}
    \sum_{\chi\chi^\prime}
    \mathcal{P}_\chi
    \Big\{
    \mathcal{I}^{\ket{q\vec{k}\uparrow,\chi}}_{\ket{q^\prime\vec{k}^\prime\!\alpha,\chi^\prime}}
    -
    \mathcal{I}^{\ket{q^\prime\vec{k}\alpha,\chi}}_{\ket{q\vec{k}^\prime\uparrow,\chi^\prime}}
                            \nonumber\\
    &\hspace*{76pt}-
    \Big[
    \mathcal{I}^{\ket{q\vec{k}\downarrow,\chi}}_{\ket{q^\prime\vec{k}^\prime\!\alpha,\chi^\prime}}
    -
    \mathcal{I}^{\ket{q^\prime\vec{k}\alpha,\chi}}_{\ket{q\vec{k}^\prime\downarrow,\chi^\prime}}
    \Big]
    \Big\}
    ,
    \end{align}
 \item
 \emph{heat current} $I_\text{Q}=\big(I_\text{Q}^L-I_\text{Q}^R\big)/2$
    \begin{align}\label{Eq:I_Q_def}
    I_\text{Q}^q
    =\ &
    \sum_{q^\prime}
    \sum_{\vec{k}\vec{k}^\prime}
    \sum_{\sigma\sigma^\prime}
    \sum_{\chi\chi^\prime}
    \mathcal{P}_\chi
    \Big\{
    \big(\varepsilon_{\vec{k}\sigma}^q-\mu_{\sigma}^q\big)
    \mathcal{I}^{\ket{q\vec{k}\sigma,\chi}}_{\ket{q^\prime\vec{k}^\prime\!\sigma^\prime,\chi^\prime}}
                            \nonumber\\
    &\hspace*{37pt}-
    \big(\varepsilon_{\vec{k}\sigma}^{q^\prime}+\Delta_{\chi\chi^\prime}-\mu_{\sigma^\prime}^q\big)
    \mathcal{I}^{\ket{q^\prime\vec{k}\sigma,\chi}}_{\ket{q\vec{k}^\prime\sigma^\prime,\chi^\prime}}
    \Big\}
    ,
    \end{align}
\end{itemize}
where $\Delta_{\chi\chi^\prime}=\varepsilon_\chi-\varepsilon_{\chi^\prime}$ and
    \begin{equation}
    \mathcal{I}^{\ket{q\vec{k}\sigma,\chi}}_{\ket{q^\prime\vec{k}^\prime\!\sigma^\prime,\chi^\prime}}
    =
    W^{\ket{q\vec{k}\sigma,\chi}}_{\ket{q^\prime\vec{k}^\prime\!\sigma^\prime,\chi^\prime}}
    f_{q\sigma}\big(\varepsilon_{\vec{k}\sigma}^q\big)
    \Big[1-f_{q^\prime\!\sigma^\prime}\big(\varepsilon_{\vec{k}^\prime\!\sigma^\prime}^{q^\prime}\big)\Big]
    .
    \end{equation}
 In Eqs.~(\ref{Eq:I_C_def})-(\ref{Eq:I_Q_def}), $\mathcal{P}_\chi$ represents the probability of finding the
 impurity in the magnetic state $\ket{\chi}$, and
    $
    f_{q\sigma}(\varepsilon)
    =
    \big\{1+\textrm{exp}\big[(\varepsilon-\mu_\sigma^q)/T_q\big]\big\}^{-1}
    $
is the Fermi-Dirac distribution function for the $q$th electrode, with $T_q$ denoting the temperature of the electrode
expressed in units of energy (i.e. $k_\text{B}\equiv 1$). Moreover, the notation for the system's complete state
$\ket{q\vec{k}\sigma,\chi}\equiv\ket{q\vec{k}\sigma}_\textrm{el}\otimes\ket{\chi}_\textrm{imp}$ is used, and the
Fermi golden rule transition rates are given by
    \begin{equation}
    W^{\ket{i}}_{\ket{j}}=
    \frac{2\pi}{\hbar}|\bra{j}\mathcal{H}_\textrm{int}\ket{i}|^2\delta(E_j-E_i),
    \end{equation}
where $\ket{i}$ and $\ket{j}$ are the initial and final states, respectively, while $E_i$ and $E_j$ denote the corresponding total energy
of the system. If, e.g., $\ket{i}=\ket{q\vec{k}\sigma,\chi}$, then
$E_i=E_{q\vec{k}\sigma,\chi}=\epsilon_{\vec{k}\sigma}^q+\mu_\sigma^q+\varepsilon_\chi$, with  $\epsilon_{\vec{k}\sigma}^q$
being the conduction electron energy measured with respect to the electrochemical potential $\mu_\sigma^q$,
$\epsilon_{\vec{k}\sigma}^q\equiv\varepsilon_{\vec{k}\sigma}^q-\mu_\sigma^q$, and
$\varepsilon_\chi$ standing for the
eigenenergy of the impurity in the state $\ket{\chi}$, see Fig.~\ref{Fig:1}(a).
Finally, the spin-dependent  electrochemical potential of the $q$th electrode can be written as
$\mu_\sigma^q=\mu_0+e\eta_q(\delta V+\eta_\sigma \delta V_\text{S})/2$,
with $\eta_{L(R)}\equiv\pm1$ and $\eta_{\uparrow(\downarrow)}=\pm1$,
together with $\delta V$  and $\delta V_\text{S}$ representing the voltage and spin bias, respectively.

It is worth of note that
Eqs.~(\ref{Eq:I_C_def}) and Eq.~(\ref{Eq:I_S_def}) are generally valid for arbitrary voltage and thermal bias.
In turn, Eq.~(\ref{Eq:I_Q_def})  for the heat current is valid in the limit of $\delta V\to 0$ and $\delta V_S\to 0$.
Moreover,  the energy factor in Eq.~(\ref{Eq:I_Q_def}) corresponds to the energy measured from the
spin-dependent electrochemical potential $\mu_\sigma^q$ of the $q$th electrode.
\footnote{For other possible definitions of a heat
current see Sec.~1.3 of Ref.~[\protect\onlinecite{Barnard_book}] or Sec. 3.9B of Ref.~[\protect\onlinecite{Mahan_book}]}
Finally, it should be noticed that Eqs.~(\ref{Eq:I_S_def})-(\ref{Eq:I_Q_def}) involve both single- ($q=q^\prime$)
and two-electrode ($q\neq q^\prime$)  electron transfer processes. In the case of energy transport, the
single-electrode processes contribute only if scattering on the impurity leads to a change in the electron energy.

In order to make use of Eqs.~(\ref{Eq:I_C_def})-(\ref{Eq:I_Q_def}) one also needs to know the probabilities of
finding the impurity in a specific magnetic state $\ket{\chi}$, which here are determined from the set of stationary master equations
    \begin{equation}\label{Eq:master_eqs}
    \underset{\chi}{\forall}\
    \sum_{\chi^\prime}
    \sum_{qq^\prime}
    \Big\{
    \mathcal{P}_{\chi^\prime}\gamma_{\chi^\prime\chi}^{qq^\prime}
    -
    \mathcal{P}_\chi\gamma_{\chi\chi^\prime}^{qq^\prime}
    \Big\}
    =
    0,
    \end{equation}
with the probability normalization condition $\sum_\chi\mathcal{P}_\chi=1$.
The golden rule  transition rate
$\gamma_{\chi\chi^\prime}^{qq^\prime}=\sum_{\vec{k}\vec{k}^\prime}\sum_{\sigma\sigma^\prime}\mathcal{I}^{\ket{q\vec{k}\sigma,\chi}}_{\ket{q^\prime\vec{k}^\prime\!\sigma^\prime,\chi^\prime}}$ between two \emph{different} spin states $\ket{\chi}$ and $\ket{\chi^\prime}$ accompanying tunneling of a single electron between the electrodes $q$ and $q^\prime$ is given by
    \begin{align}\label{Eq:trans_rates_general}
    &
    \hspace*{-3pt}
    \gamma_{\chi\chi^\prime}^{qq^\prime}
    =
    \frac{2\pi}{\hbar}K^2
    \big(\alpha_\text{ex}^{qq^\prime}\big)^{\!2}
    \sum_{\sigma\sigma^\prime}
    \Phi_{\sigma\sigma^\prime}^{(0)qq^\prime}\!\!(\Delta_{\chi\chi^\prime})
                \nonumber\\
    &\hspace*{5pt}
    \times
    \Big\{
    \!
    \delta_{\sigma^\prime\overline{\sigma}}
    \Big[
    \delta_{\sigma\downarrow}
    \big|\mathbb{S}_{\chi^\prime\chi}^-\big|^2
    +
    \delta_{\sigma\uparrow}
    \big|\mathbb{S}_{\chi^\prime\chi}^+\big|^2
    \Big]
    +
    \delta_{\sigma^\prime\sigma}
    \big|\mathbb{S}_{\chi^\prime\chi}^z\big|^2
    \Big\}
    ,
    \end{align}
where $\alpha_\text{ex}^{qq^\prime}\equiv\alpha_\text{ex}\nu_q\nu_{q^\prime}$,
$\mathbb{S}_{\chi^\prime\chi}^k\equiv\bra{\chi^\prime}S_k\ket{\chi}$ for $k=z,\pm$,  and
    \begin{align}\label{Eq:Phi_function_def}
    \hspace*{-2pt}
    \Phi_{\sigma\sigma^\prime}^{(n)qq^\prime}\!\!(\Delta_{\chi\chi^\prime})
    =
    \int\!\!\text{d}\omega
    \rho_\sigma^q(\omega)\rho_{\sigma^\prime}^{q^\prime}(\omega+\Delta_{\chi\chi^\prime})
    (\omega-\mu_0)^n
    &
                    \nonumber\\
    \times
    f_{q\sigma}(\omega)\Big[1-f_{q^\prime\!\sigma^\prime}(\omega+\Delta_{\chi\chi^\prime})\Big]
    &
    .
    \hspace*{-3pt}
    \end{align}

We remind that Eqs~(\ref{Eq:I_C_def})-(\ref{Eq:I_S_def}) are valid for arbitrary $T$ and also in the nonlinear regime, while Eq.~(\ref{Eq:I_Q_def}) is valid for
$\delta V\to 0$ and $\delta V_S\to 0$, with no restriction on $\delta T$. Below we will linearize these equations with respect to all variables, i.e.
 with respect to $\delta V$,  $\delta V_S$, and $\delta T$. Apart from this, we restrict our considerations to the regime of low $T$, $T\ll (2S-1)D$. However, the latter
 restriction is essential only in the anisotropic case, $D>0$, and does not have to be fulfilled for isotropic case ($D=E=0$) and for the case of no impurity in the barrier.

\subsection{Linear response regime -- kinetic coefficients}

 In the regime of linear response with respect to the voltage $\delta V$, spin voltage $\delta V_\text{S}$,
 and thermal bias $\delta T$, the formulae for charge ($I_\text{C}$), spin ($I_\text{S}$) and heat ($I_\text{Q}$)
 currents can be written in the following general form:~\cite{Barnard_book,Mahan_book}
    \begin{equation}\label{Eq:I_n_matrix_kin_coeff}
    \renewcommand{\arraystretch}{1.5}
        \begin{pmatrix}
        I_\text{C}
        \\
        I_\text{S}
        \\
        I_\text{Q}
        \end{pmatrix}
    \!
    =
    \!
        \begin{pmatrix}
        e^2\mathcal{L}_{00} & e^2\mathcal{L}_{01} & e\mathcal{L}_{02}/T
        \\
        e\hbar\mathcal{L}_{10}/2 & e\hbar\mathcal{L}_{11}/2 & \hbar\mathcal{L}_{12}/(2T)
        \\
        e\mathcal{L}_{20} & e\mathcal{L}_{21} & \mathcal{L}_{22}/T
        \end{pmatrix}
        \!\!\!
        \begin{pmatrix}
        \delta V
        \\
        \delta V_\text{S}
        \\
        \delta T
    \end{pmatrix}
    \!\!
        ,
    \end{equation}
where $\mathcal{L}_{nk}$ are the relevant \emph{kinetic coefficients} that satisfy the Onsager
relation,\cite{Onsager_Phys.Rev.37/1931_I,*Onsager_Phys.Rev.38/1931_II,deGroot_book} $\mathcal{L}_{nk}=\mathcal{L}_{kn}$.
Interestingly enough, by assuming additionally $T\ll (2S-1)D$ (for $D>0$), then only the impurity's ground state doublet $\ket{\chi_{\pm S}}$
plays a role, as the transitions to excited spin states are energetically forbidden.
Nevertheless, due to the transverse magnetic anisotropy, spin-flip scattering processes within this ground doublet are
still allowed.

Explicit form of the kinetic coefficients  in Eq.~(\ref{Eq:I_n_matrix_kin_coeff}) can be obtained by
linearization of the expressions (\ref{Eq:I_C_def})-(\ref{Eq:I_Q_def})  for  the currents, $I_\text{C}$, $I_\text{S}$,
and $I_\text{Q}$ (for a detailed derivation see App.~\ref{App:kin_coeff}). For convenience, we write these coefficients in the form:
    \begin{equation}\label{Eq:kin_coeff_matrix}
    \renewcommand{\arraystretch}{1.75}
    \hspace*{-4pt}
    \bm{\mathcal{L}}
    =    \begin{pmatrix}
        \mathcal{L}_{00} &\mathcal{L}_{01} & \mathcal{L}_{02}
        \\
        \mathcal{L}_{10}& \mathcal{L}_{11} & \mathcal{L}_{12}
          \\
        \mathcal{L}_{20} & \mathcal{L}_{21}& \mathcal{L}_{22}
    \end{pmatrix}
    \equiv
    \begin{pmatrix}
        \mathcal{L}_{00}^{(c)}&\mathcal{L}_{01}^{(s)} & \mathcal{L}_{02}^{(c)}
        \\
        \mathcal{L}_{10}^{(s)}& \mathcal{L}_{11}^{(ss)} & \mathcal{L}_{12}^{(s)}
        \\
        \mathcal{L}_{20}^{(c)} & \mathcal{L}_{21}^{(s)}& \mathcal{L}_{22}^{(c)}
    \end{pmatrix}
    \!\!
    ,
    \hspace*{-3pt}
    \end{equation}
with the elements of the matrix given by
    \begin{equation}\label{Eq:L_elements}
    \left\{
        \begin{aligned}
        &\mathcal{L}_{nk}^{(c)}
        =
        \sum_\sigma\mathcal{L}_{nk}^{\sigma}+\mathcal{L}_{nk,\downarrow\uparrow}^{(c)},
                    \\
        &\mathcal{L}_{nk}^{(s)}
        =
        \sum_\sigma\eta_\sigma\mathcal{L}_{nk}^{\sigma}+\mathcal{L}_{nk,\downarrow\uparrow}^{(s)},
                    \\
        &\mathcal{L}_{nk}^{(ss)}
        =
        \sum_\sigma\mathcal{L}_{nk}^{\sigma}+\mathcal{L}_{\downarrow\uparrow}^{(ss)},
        \end{aligned}
    \right.
    \end{equation}
\begin{table}
\setlength{\tabcolsep}{8pt}
\caption{\label{Tab:Theta_coeff}Explicit expressions for the auxiliary coefficients $\vartheta_n$, where
$\Lambda_\pm=\big|\mathbb{S}_{\chi_{-S}\chi_S}^+\big|^2\pm\big|\mathbb{S}_{\chi_{-S}\chi_S}^-\big|^2$.}
\begin{ruledtabular}
\begin{tabular}{c|cc}
\multicolumn{1}{c|}{\multirow{2}{*}{$n$}} & \multicolumn{2}{c}{$\vartheta_n$}\\[-2pt]
 & \footnotesize \emph{Isotropic} spin impurity & \footnotesize\emph{Anisotropic} spin impurity \\[4pt]\hline
\rule{0pt}{18pt}1 & $\dfrac{2}{2S+1}$ & 1\\
\rule{0pt}{18pt}2 & $\dfrac{4}{3}S(S+1)$ & $\dfrac{\Lambda_-^2}{\Lambda_+}$\\
\rule{0pt}{18pt}3 & $\dfrac{4}{3}S(S+1)$ & $\Lambda_+$\\[7pt]
\end{tabular}
\end{ruledtabular}
\end{table}
and
    \begin{align}
    \hspace*{-3pt}
    \mathcal{L}_{nk}^{\sigma}
    =
    \frac{\pi}{\hbar}
    \frac{K^2}{T}
    \vartheta_1
    \!
    \sum_{\chi}
    \!
    \Big[
    \alpha_\text{d}
    +
    \eta_\sigma
    \alpha_\text{ex}^{LR}\,
    \mathbb{S}_{\chi\chi}^z
    \Big]^2
    \!
    \mathcal{F}_{\sigma\sigma}^{(\delta_{n2}+\delta_{k2})LR}
     ,
     \hspace*{-2pt}
    \end{align}
    \begin{multline}\label{Eq:L_c}
    \mathcal{L}_{nk,\downarrow\uparrow}^{(c)}
    =
    \frac{\pi}{\hbar}
    \frac{K^2}{T}
    \Bigg\{
    \vartheta_3
    \big(\alpha_\text{ex}^{LR}\big)^{\!2}
    \sum_\sigma
    \mathcal{F}_{\sigma\overline{\sigma}}^{(\delta_{n2}+\delta_{k2})LR}
                            \\
    -
    \vartheta_2
    \Omega
    \big(\alpha_\text{ex}^{LR}\big)^{\!4}
    \sum_{\sigma\sigma^\prime}
    \eta_\sigma
    \eta_{\sigma^\prime}
     \mathcal{F}_{\sigma\overline{\sigma}}^{(\delta_{n2})LR}
     \mathcal{F}_{\sigma^\prime\overline{\sigma^\prime}}^{(\delta_{k2})LR}
     \Bigg\}
     ,
    \end{multline}
    \begin{multline}\label{Eq:L_s}
    \mathcal{L}_{nk,\downarrow\uparrow}^{(s)}
    =
    -
    \frac{\pi}{\hbar}
    \frac{K^2}{T}
    \vartheta_2
    \Omega
     \big(\alpha_\text{ex}^{LR}\big)^{\!2}
                            \\
    \times\!
    \sum_{q\sigma}
    \eta_q
    \eta_\sigma
    \big(\alpha_\text{ex}^{qq}\big)^{\!2}
     \mathcal{F}_{\sigma\overline{\sigma}}^{(\delta_{n2}+\delta_{k2})LR}
     \mathcal{F}_{\uparrow\downarrow}^{(0)qq}
     ,
    \end{multline}
    \begin{multline}\label{Eq:L_ss}
    \mathcal{L}_{\downarrow\uparrow}^{(ss)}
     =
    \frac{\pi}{\hbar}
    \frac{K^2}{T}
    \Bigg\{
    \vartheta_3
    \sum_q
     \big(\alpha_\text{ex}^{qq}\big)^{\!2}
    \mathcal{F}_{\uparrow\downarrow}^{(0)qq}
                    \\
    -
    \vartheta_2
    \Omega
    \sum_{qq^\prime}
    \eta_q
    \eta_{q^\prime}
     \big(\alpha_\text{ex}^{qq}\big)^{\!2}
    \big(\alpha_\text{ex}^{q^\prime q^\prime}\big)^{\!2}
     \mathcal{F}_{\uparrow\downarrow}^{(0)qq}
     \mathcal{F}_{\uparrow\downarrow}^{(0)q^\prime\! q^\prime}
     \Bigg\}
     ,
    \end{multline}
where $\mathcal{F}_{\sigma\sigma^\prime}^{(n)qq^\prime}=T\phi_{\sigma\sigma^\prime}^{(n,0)qq^\prime}=
\Phi_{\sigma\sigma^\prime}^{(n)qq^\prime}\!\!(0)\big|_\text{eq}$, see Eqs.~(\ref{Eq:Phi_function_def}) and~(\ref{Eq:aux_phi_function}),
and the subscript `$\text{eq}$' means the quantity to be taken at $\delta V=\delta V_\text{S}=\delta T = 0$.
Furthermore, in the above equations $\Omega$ is defined as
    \begin{equation}
    \Omega
    =
   \Bigg[
    \sum_{qq^\prime}
    \!\!
    \big(\alpha_\text{ex}^{qq^\prime}\big)^2\,
    \Phi_{\uparrow\downarrow}^{(0)qq^\prime}\!\!(0)\big|_\text{eq}
    \Bigg]^{-1}
    ,
    \end{equation}
while $\vartheta_n$ ($n=1,2,3$) is defined in Tab.~\ref{Tab:Theta_coeff} for isotropic and anisotropic spins.
We remind that in the \emph{anisotropic} case ($D\neq0$ and $E\neq0$) only the two degenerate states of
lowest energy are included in the sums over $\chi$ due to the condition $T\ll (2S-1)D$, whereas in the \emph{isotropic} case ($D=E=0$)
all states are taken into account as they all are degenerate and the condition $T\ll (2S-1)D$ is irrelevant.

The kinetic coefficients consist of two terms, see Eqs.~(\ref{Eq:L_elements}). The first term originates from
electron tunneling with conserved electron spin, while the second term  in each coefficient takes into account
tunneling associated with reversal of electron spin (and thus also with a change in magnetic state of the impurity).
Note that for an anisotropic spin impurity with vanishing transverse magnetic anisotropy, $E\rightarrow0$ while $D>0$,
one finds  $\mathbb{S}_{\chi_{-S}\chi_S}^\pm=0$  and thus only the components $\mathcal{L}_{nk}^{\sigma}$ of the
kinetic coefficients in Eqs.~(\ref{Eq:L_elements}) survive,
whereas the terms given by  Eqs.~(\ref{Eq:L_c})-(\ref{Eq:L_ss}) turn to zero.

To find numerical values of the kinetic coefficients $\mathcal{L}_{nk}$,  we need to calculate all the factors
of the type $\mathcal{F}_{\sigma\sigma^\prime}^{(n)qq^\prime}$. The key problem is that this requires evaluation
of energy integrals involving DOS of electrodes, which in general can be an arbitrary function of energy.
Taking into account  the fact that  transport properties at low temperature and in the linear response regime are
determined by the electrodes' DOS in the vicinity of the equilibrium electrochemical potential $\mu_0$, we expand
the spin-dependent DOS of the $q$th electrode into a series,
    \begin{equation}
    \rho_\sigma^q(\omega)
    =
    \sum_k
    \frac{\big[\rho_\sigma^q(\mu_0)\big]^{(k)}}{k!}
    (\omega-\mu_0)^k
    \end{equation}
with
    \begin{equation}
    \big[\rho_\sigma^q(\mu_0)\big]^{(k)}
    =
    \frac{\partial^k \rho_\sigma^q(\omega)}{\partial\omega^k}\big|_{\omega =\mu_0}
    .
    \end{equation}
This, in turn, allows for calculating the energy integrals in question,
    \begin{align}\label{Eq:Phi_function_eq_int}
    &
    \hspace*{-18pt}
    \mathcal{F}_{\sigma\sigma^\prime}^{(n)qq^\prime}
    =
    \sum_{kl}
    \frac{\big[\rho_\sigma^q(\mu_0)\big]^{(k)}}{k!}
    \frac{\big[\rho_{\sigma^\prime}^{q^\prime}(\mu_0)\big]^{(l)}}{l!}
                \nonumber\\
    &\hspace*{55pt}
    \times
    \!\!
    \int\!\text{d}\omega
    \big(\omega-\mu_0\big)^{n+k+l}
    f(\omega)\Big[1-f(\omega)\Big]
                \nonumber\\
    =\ &
    \sum_{kl}
    \frac{\big[\rho_\sigma^q(\mu_0)\big]^{(k)}}{k!}
    \frac{\big[\rho_{\sigma^\prime}^{q^\prime}(\mu_0)\big]^{(l)}}{l!}
    \Theta_{n+k+l}T^{n+k+l+1}
    ,
    \end{align}
where
    \begin{equation}\label{Eq:Theta_n}
    \Theta_n
    =
    (-1)^{n/2+1}(1-2^{1-n})(2\pi)^n\text{B}_n,
    \end{equation}
and $\text{B}_n$ stands for the Bernoulli number. It can be noticed that $\Theta_n=0$ if $n$ is an odd number,
while first several even terms are: $\Theta_0=1$, $\Theta_2=\pi^2/3$, $\Theta_4=7\pi^4/15$, $\Theta_6=31\pi^6/21$, etc.

Having found all the kinetic coefficients, one can calculate the experimentally measurable coefficients discussed in
Sec.~\ref{Sec:Background_thermoel}, which are directly related to the kinetic coefficients.
In particular,  Eqs.~(\ref{Eq:G_def})-(\ref{Eq:S_S_def})  can be expressed in terms of the kinetic coefficients $\mathcal{L}_{nk}$ as
\begin{itemize}[leftmargin=*]
\item
Conductances:
    \begin{equation}
    \left\{
    \begin{aligned}
    &G = e^2\mathcal{L}_{00}
   =
   G_\uparrow+G_\downarrow+G_{\downarrow\uparrow}^{(c)},
                    \\
   &G^\text{m}=e^2\mathcal{L}_{01}
   =
   G_\uparrow-G_\downarrow+G_{\downarrow\uparrow}^{(s)},
                    \\
   &G_\text{S}^\text{m}=\frac{e\hbar}{2}\mathcal{L}_{10}
   =
   \frac{\hbar}{2e}\Big[G_\uparrow-G_\downarrow+G_{\downarrow\uparrow}^{(s)}\Big],
                    \\
   & G_\text{S}=\frac{e\hbar}{2}\mathcal{L}_{11}
   =
   \frac{\hbar}{2e}\Big[G_\uparrow+G_\downarrow+G_{\downarrow\uparrow}^{(ss)}\Big],
   \end{aligned}
   \right.
    \end{equation}
where $G^\text{m}$ and $G^\text{m}_\text{S}$ are related as $G^\text{m}_\text{S} =(\hbar /2e)G^\text{m}$.
Above, $G_\sigma=e^2\mathcal{L}_{00}^\sigma$
is the electric conductance of the spin-$\sigma$ channel due to spin conserving  electron tunneling between the left and right electrodes,
whereas $G_{\downarrow\uparrow}^{(c)}=e^2\mathcal{L}_{00,\downarrow\uparrow}^{(c)}$ and
$G_{\downarrow\uparrow}^{(s/ss)}=e^2\mathcal{L}_{\downarrow\uparrow}^{(s/ss)}$ represent a
contribution to conductance stemming from tunneling with spin-flip processes.  Note that for the $(ss)$-component
we have only single-electrode processes. Such processes modify spin  state of the molecule without transferring
any charge across the junction, or in other words, they transfer spin without transferring charge.
\item
Peltier coefficients:  using the notation $\Pi\equiv\Pi_0$ and $\Pi_\text{S}\equiv\Pi_1$, one can write
    \begin{equation}
    \hspace*{-2pt}
    \Pi_n
    =
    \left[-\frac{1}{|e|}\right]^{\delta_{n0}}
    \!
    \left[\frac{2}{\hbar}\right]^{\delta_{n1}}
    \!
    \frac{
    \mathcal{L}_{2n}
    }{
    \mathcal{L}_{nn}
    }
    \quad
    \text{for}
    \quad
    n=0,1
    ,
    \end{equation}
\item
Thermal conductance:
    \begin{equation}
     \kappa
    =
    \frac{1}{T}
    \Bigg[
    \mathcal{L}_{22}-\frac{\big(\mathcal{L}_{02}\big)^2}{\mathcal{L}_{00}}
    \Bigg]
    ,
    \end{equation}
\item
Thermopowers: using the notation $\mathcal{S}\equiv \mathcal{S}_0$  and $\mathcal{S}_\text{S}\equiv \mathcal{S}_1$, one finds
    \begin{equation}
    \mathcal{S}_n=-\frac{1}{|e|T}
    \frac{
    \mathcal{L}_{n2}
    }{
    \mathcal{L}_{nn}
    }
     \quad
    \text{for}
    \quad
    n=0,1
    .
    \end{equation}
\end{itemize}

\section{\label{Sec:Results}Results and discussion}

As already mentioned in the Introduction,  thermoelectric effects become revealed when DOS is energy dependent around the
Fermi level and there is a particle-hole asymmetry. In the conceptually simplest case assumed here,
DOS in one electrode is constant on the energy scale of interest, while DOS of the other electrode is a linear
function of energy. For this reason, the DOS of the left electrode in the vicinity of the Fermi level is assumed to be constant and \emph{spin-dependent},
    \begin{equation}\label{Eq:rho_L_approx}
    \rho_\sigma^L(\omega)
    \approx
    \rho_\sigma^L(\mu_0)
    =
    \frac{\rho^L}{2}
    \big[
    1+\eta_\sigma P_L
    \big]
    ,
    \end{equation}
whereas the right electrode is assumed to be \emph{nonmagnetic} with the DOS linearly dependent on energy around the Fermi level,
    \begin{align}\label{Eq:rho_R_approx}
    \rho_\sigma^R(\omega)
    \approx\
    &
    \rho_\sigma^R(\mu_0)
    +
    \big[\rho_\sigma^R(\mu_0)\big]^{(1)}(\omega-\mu_0)
                \nonumber\\
    =\
    &
    \frac{\rho^R}{2}
    \big[
    1+x_R(\omega-\mu_0)
    \big]
    ,
    \end{align}
with  $x_R\equiv \big[\rho^R(\mu_0)\big]^{(1)}/\rho^R$, where we took into account that
$\big[\rho_\uparrow^R(\mu_0)\big]^{(1)}=\big[\rho_\downarrow^R(\mu_0)\big]^{(1)}=\big[\rho^R(\mu_0)\big]^{(1)}/2$.
In the above equations $\rho_L$ and $\rho_R$ denote the total DOS at the Fermi level in the left and right electrodes, respectively.
The above approximations correspond to a situation, when the left
electrode has a relatively flat DOS around the Fermi level, while the right electrode
is characterized by DOS with a steep slope at the Fermi level.
Although the coefficient  $x_R$ can in general  be both positive and negative, the DOS has to be a non-negative
function of energy, $\rho_\sigma^R(\omega)\geqslant0$. This imposes some restrictions on the energy range where this approximation is applicable.
Moreover, since the electrons and holes are distributed around the Fermi level in the energy window of the order of $T$, this imposes also the following
condition on the temperature: $x_RT\ll1$.

Taking the above into account, one finds
    \begin{multline}
    \mathcal{F}_{\sigma\sigma^\prime}^{(n)LR}
    =
    \frac{1}{4}
    \rho^R
    \rho^L
    \big[
    1+\eta_\sigma P_L
    \big]
                \\
    \times
    \Big\{
    \Theta_{n}T^{n+1}
    +
    x_R
    \Theta_{n+1}T^{n+2}
    \Big\}
    .
    \end{multline}
Since we have assumed that the right electrode is nonmagnetic, the second spin index in
$\mathcal{F}_{\sigma\sigma^\prime}^{(n)LR}$ plays in fact no role,
and thus we omit it henceforth,
    \begin{equation}
   \left\{
        \begin{aligned}
        &
        \mathcal{F}_{\sigma}^{(0)LR}
        =
        \frac{1}{4}
        \rho^L\rho^R\big[1+\eta_\sigma P_L\big]T,
                    \\
        &
       \mathcal{F}_{\sigma}^{(1)LR}
       =
       \frac{\pi^2}{12}\rho^L\rho^Rx_R\big[1+\eta_\sigma P_L\big]T^3,
                    \\
       &
        \mathcal{F}_{\sigma}^{(2)LR}
        =
        \frac{\pi^2}{12}
        \rho^L\rho^R\big[1+\eta_\sigma P_L\big]T^3.
        \end{aligned}
   \right.
   \end{equation}
Analogous expressions can be derived for single-electrode $\mathcal{F}$-functions,
    \begin{equation}
    \left\{
        \begin{aligned}
        &\mathcal{F}_{\uparrow\downarrow}^{(0)LL}
        =
        \frac{\big(\rho^L\big)^2}{4}
        \big[
        1-P_L^2
        \big]
        T,
                \\
        &\mathcal{F}_{\uparrow\downarrow}^{(0)RR}
        =
        \frac{\big(\rho^R\big)^2}{4}
        \Big[
        1
        +
        \frac{\pi^2}{3}
        x_R^2T^2
        \Big]T
        .
        \end{aligned}
        \right.
    \end{equation}

In consequence, the above assumptions allow to write the matrix $\bm{\mathcal{L}}$ in the form
    \begin{equation}\label{Eq:kin_coeff_matrix_final}
    \renewcommand{\arraystretch}{2.25}
    \hspace*{-4pt}
    \bm{\mathcal{L}}
    =
    \begin{pmatrix}
        \mathcal{L}_{0}&\mathcal{L}_s & \dfrac{\pi^2}{3}x_RT^2\mathcal{L}_{0}
        \\
        \mathcal{L}_s& \mathcal{L}_{1} & \dfrac{\pi^2}{3}x_RT^2\mathcal{L}_{s}
        \\
        \dfrac{\pi^2}{3}x_RT^2\mathcal{L}_{0} & \dfrac{\pi^2}{3}x_RT^2\mathcal{L}_{s}& \mathcal{L}_{2}
    \end{pmatrix}
    \!\!
    ,
    \hspace*{-3pt}
    \end{equation}
with $\mathcal{L}_n$ (for $n=0,1,2,s$) having the form
    \begin{equation}\label{Eq:L_final}
    \mathcal{L}_{n}
    =
    \Gamma\lambda
    P_L^{\delta_{ns}}
    \Big[
    \frac{\pi^2}{3}T^2
    \Big]^{\delta_{n2}}
    \Big[
    \mathcal{T}_\text{sc}
    +
    \mathcal{T}_\text{sf}^{(n)}
    \Big]
    ,
    \end{equation}
where $\Gamma=\pi K^2\rho^2/\hbar$ with  $\rho\equiv\rho^L$, and  $\lambda=\rho^R/\rho^L$. Furthermore,
    \begin{equation}
    \mathcal{T}_\text{sc}
    \equiv
    \big(
    \alpha_\text{d}
    \big)^{\!2}
    +
    \frac{\vartheta_1}{2}
    \big(
    \alpha_\text{ex}^{LR}
    \big)^{\!2}
    \sum_{\chi}
    \!
    \big(
    \mathbb{S}_{\chi\chi}^z
    \big)^{\!2}
    \end{equation}
represents the \emph{spin-conserving} part of the kinetic coefficients. The first term of $\mathcal{T}_\text{sc}$ corresponds to the direct
tunneling of electrons between the electrodes, whereas the second term accounts for
two-electrode tunneling processes during which electrons
traversing the barrier interact {\it via} exchange coupling with the impurity, keeping, however, their spin orientation unchanged.
On the other hand, the \emph{spin-flip} part
$\mathcal{T}_\text{sf}^{(n)}$ stands for all tunneling processes (including both single- and two-electrode ones) in which
the spin of an electron becomes reversed due to scattering on the spin impurity,
    \begin{equation}
    \mathcal{T}_\text{sf}^{(0)}
    =
    \big(\alpha_\text{ex}^{LR}\big)^{\!2}
    \Bigg\{
    \frac{\vartheta_3}{2}
    -
    \frac{
    \vartheta_2
    P_L^2
    \big(\alpha_\text{ex}^{LR}\big)^{\!2}
    }{
    \big(\widetilde{\alpha}_\text{ex}\big)^{\!2}
    }
    \Bigg\}
     ,
    \end{equation}
    \begin{multline}\label{Eq:T_1_final}
    \mathcal{T}_\text{sf}^{(1)}
    =
    \big(\alpha_\text{ex}^{LR}\big)^{\!2}
       \frac{
    \vartheta_3
    \big(\alpha_\text{ex}^+\big)^{\!2}
    }{
    2
    \big(\widetilde{\alpha}_\text{ex}\big)^{\!2}
    }
                                \\
    +
    \frac{
    1
    }{
    4
    \big(\widetilde{\alpha}_\text{ex}\big)^{\!2}
    }
    \Bigg\{
    \vartheta_3
    \big(\alpha_\text{ex}^+\big)^{\!4}
    -
    \vartheta_2
    \big(
    \alpha_\text{ex}^-
     \big)^{\!4}
    \Bigg\}
     ,
    \end{multline}
    \begin{equation}
    \mathcal{T}_\text{sf}^{(2)}
    =
    \big(\alpha_\text{ex}^{LR}\big)^{\!2}
    \Bigg\{
    \frac{\vartheta_3}{2}
    -
    \frac{\pi^2}{3}
    x_R^2
    T^2
    \cdot
    \frac{
    \vartheta_2
    P_L^2
    \big(\alpha_\text{ex}^{LR}\big)^{\!2}
    }{
    \big(\widetilde{\alpha}_\text{ex}\big)^{\!2}
    }
    \Bigg\}
     ,
     \hspace*{-3pt}
    \end{equation}
    \begin{equation}\label{Eq:T_s_final}
    \mathcal{T}_\text{sf}^{(s)}
    =
     -
     \big(\alpha_\text{ex}^{LR}\big)^{\!2}
     \cdot
    \frac{
    \vartheta_2
     \big(
    \alpha_\text{ex}^-
    \big)^{\!2}
    }{
    2
    \big(\widetilde{\alpha}_\text{ex}\big)^{\!2}
    } ,
    \end{equation}
where
    \begin{equation}
    \big(\widetilde{\alpha}_\text{ex}\big)^{\!2}
    =
    2
    \big(\alpha_\text{ex}^{LR}\big)^{\!2}
    +
    \big(\alpha_\text{ex}^+\big)^{\!2}
    \end{equation}
and
    \begin{equation}
    \hspace*{-5pt}
    \big(\alpha_\text{ex}^\pm\big)^{\!2}
    =
    \frac{1}{\lambda}
    \big(\alpha_\text{ex}^{LL}\big)^{\!2}
    \big[
    1-P_L^2
    \big]
    \pm
    \lambda
    \big(\alpha_\text{ex}^{RR}\big)^{\!2}
    \Big[
    1
    +
    \frac{\pi^2}{3}
    x_R^2T^2
    \Big]
    .
    \hspace*{-2pt}
    \end{equation}
When deriving the above formulae, we also took into account that $\Omega
    =4/\lambda T\rho^2 \big(\widetilde{\alpha}_\text{ex}\big)^{\!2}$.
One can notice that the effective coefficients $\big(\alpha_\text{ex}^\pm\big)^{\!2}$ have a clear physical
meaning, namely they represent a contribution from single-electrode spin-exchange tunneling processes. In turn,
such processes involving two different electrodes (i.e. for electrons traversing the junction)
are described by $\big(\alpha_\text{ex}^{LR}\big)^{\!2}$. Interestingly enough, it can be immediately seen
that, unlike other coefficients,  $\mathcal{T}_\text{sf}^{(1)}$  apart from the part corresponding  to
tunneling of electrons between the left and right electrodes, includes also the term  originating from
single-electrode tunneling processes. The physical notion of this observation will be discussed in detail in further sections.

Employing the above form of the $\bm{\mathcal{L}}$-matrix, Eq.~(\ref{Eq:kin_coeff_matrix_final}), one obtains:
\begin{itemize}[leftmargin=*]
\item
Conductances:
    \begin{equation}\label{Eq:G_S_def}
    \renewcommand{\arraystretch}{2}
    \bm{G}
    =
        \begin{pmatrix}
        e^2\mathcal{L}_0
        &
        e^2\mathcal{L}_s
                    \\
        \dfrac{e\hbar}{2}\mathcal{L}_s
        &
        \dfrac{e\hbar}{2}\mathcal{L}_1
        \end{pmatrix}
        ,
   \end{equation}
\item
Peltier coefficients:
    \begin{equation}\label{Eq:Pi_final}
    \Pi
    =
    -
    \frac{\pi^2}{3|e|}
    x_RT^2
    \quad
    \text{and}
    \quad
    \Pi_\text{S}
    =
    -
    \frac{2|e|}{\hbar}
    \frac{\mathcal{L}_s}{\mathcal{L}_1}
    \Pi
    ,
    \end{equation}
\item
Thermal conductance:
    \begin{equation}
     \kappa
    =
    \frac{1}{T}
    \Bigg[
    \mathcal{L}_{2}
    -
   \Big(\frac{\pi^2}{3}x_RT^2\Big)^{\!2}
    \mathcal{L}_{0}
    \Bigg]
    ,
    \end{equation}
\item
Thermopowers:
    \begin{equation}\label{Eq:S_final}
    \mathcal{S}
    =
    -
    \frac{\pi^2}{3|e|}
    x_RT
    \quad
    \text{and}
    \quad
    \mathcal{S}_\text{S}
    =
    \frac{\mathcal{L}_s}{\mathcal{L}_1}
    \mathcal{S}
    =
    \frac{G_\text{S}^\text{m}}{G_\text{S}}
    \mathcal{S}
    ,
    \end{equation}
\item
Figures of merit:
    \begin{equation}\label{Eq:ZT_final}
    \text{ZT}
    =
    \frac{
    \mathcal{L}_0
    \Big(
    \frac{\pi^2}{3}
    \Big)^2
    x_R^2
    T^4
    }{
    \mathcal{L}_{2}
    -
    \mathcal{L}_{0}
    \Big(
    \frac{\pi^2}{3}
    \Big)^2
    x_R^2
    T^4
    }
    \end{equation}
and
    \begin{equation}\label{Eq:ZT_S_final}
    \text{ZT}_\text{S}
    =
    \frac{
    \mathcal{L}_{s}^2
    }{
    \mathcal{L}_{1}\mathcal{L}_{0}
    }
    \text{ZT}
    =
    -\frac{2|e|}{\hbar}
    \frac{
    \big(G_\text{S}^\text{m}\big)^{\!2}
    }{
    GG_\text{S}
    }
    \text{ZT}
    .
    \end{equation}
\end{itemize}

Combining Eqs.~(\ref{Eq:Pi_final}) and~(\ref{Eq:S_final}), one straightforwardly gets the \emph{Thompson's second
relation}~\cite{deGroot_book} and its spin analog, which in general establish a connection between the respective
Peltier and Seebeck effects,
    \begin{equation}\label{Eq:Thompson_relation}
    \Pi=\mathcal{S}T
    \quad
    \text{and}
    \quad
    \Pi_\text{S}=
    -\frac{2|e|}{\hbar}
    \mathcal{S}_\text{S}T
    .
    \end{equation}
Since the conventional Peltier coefficient $\Pi$ and thermopower $\mathcal{S}$ are independent of the specific properties of the
spin impurity, such as a spin number and values of uniaxial and transverse magnetic anisotropy constants, in the
following discussion we focus exclusively on its spin-dependent counterparts. In addition, because spin-dependent
Peltier coefficient $\Pi_\text{S}$ is generally related  to a spin-dependent thermopower
$\mathcal{S}_\text{S}$ \emph{via} Eq.~(\ref{Eq:Thompson_relation}), we focus mainly only the later one.

\subsection{\label{Sec:NO_impurity}Absence of magnetic impurity}
The above formulas are general in the sense that they apply to anisotropic and  isotropic impurities.
Before considering these two situations, it is instructive to analyze first the simplest case of
a magnetic tunnel junction (MTJ)
without spin impurity in the barrier, which essentially corresponds to setting $\alpha_\text{ex}=0$ in the above formulas.
In such a case the kinetic coefficients take a simple form,
    \begin{equation}\label{Eq:L_MTJ}
    \mathcal{L}_0
    =
    \mathcal{L}_1
    =
    \frac{1}{P_L}
    \mathcal{L}_s
    =
    \frac{3}{\pi^2T^2}
    \mathcal{L}_2
    =\Gamma\lambda\big(\alpha_\text{d}\big)^2.
    \end{equation}

Assuming $\alpha_\text{d}=1$,  one can write the conductance matrix for MTJ as follows:
    \begin{equation}\label{Eq:G_MTJ}
    \renewcommand{\arraystretch}{1.75}
    \bm{G}^\text{MTJ}
    =
    \Gamma\lambda
        \begin{pmatrix}
        e^2
        &
        e^2
        P_L
                    \\
        \dfrac{e\hbar}{2}
        P_L
        &
        \dfrac{e\hbar}{2}
        \end{pmatrix}
        .
   \end{equation}
Since $\Gamma \lambda = \pi K^2\rho^L\rho^R/\hbar $, the conductance matrix $\bm{G}^\text{MTJ}$  in
the model under consideration is proportional to the product of the DOS at the
Fermi level in the two electrodes and to the square of the direct tunneling parameter $T_\text{d}\equiv K$. In turn,
the thermal conductance is given by
    \begin{align}\label{Eq:kappa_MTJ}
     \kappa^\text{MTJ}
    &
    =
    \frac{\pi^2}{3}
    \Gamma
    \lambda T
    \Big[
    1
    -
    \frac{\pi^2}{3}x_R^2T^2
    \Big]
                \nonumber\\
    &=
    L_0
    G^\text{MTJ}T
    \Big[
    1
    +
    |e| x_R \Pi
    \Big]
    ,
    \end{align}
where the first term represents the Wiedemann-Franz (WF) law  that relates the thermal and electrical
conductances as $\kappa=L_0GT$, with $L_0=\pi^2/(3e^2)$ being the Lorentz number (recall that in
this paper we set  $k_\text{B}\equiv1$).  The WF law applies for instance to transport in Fermi liquid
bulk metals, but it generally breaks down in nanoscopic
systems,~\cite{Kubala_Phys.Rev.Lett.100/2008,Tsaousidou_J.Phys.:Condens.Matter22/2010} though it can be recovered
in the situation when the system reaches an effective Fermi liquid state, e.g., as in strongly correlated quantum
dots when the Kondo effect occurs.~\cite{Boese_Europhys.Lett.56/2001,Dong_J.Phys.:Condens.Matter14/2002,Krawiec_Phys.Rev.B73/2006}
Due to the presence of the second term in Eq.~(\ref{Eq:kappa_MTJ}), the WF law in the current situation is generally violated.
However, this deviation is rather small in the applicability range of the model,~$x_RT\ll1$.

As mentioned above, the thermopower $\mathcal{S}$ is independent of the presence of the impurity, and is given by the formula~(\ref{Eq:S_final}).
It is also worth noting that the formula for thermopower $\mathcal{S}^\text{MTJ}$, Eq.~(\ref{Eq:S_final}),
obeys the Mott's formula~\cite{Sivan_Phys.Rev.B33/1986,Lunde_J.Phys.:Condens.Matter17/2005}
    \begin{equation}\label{Eq:S_Mott}
    \mathcal{S}^\text{MTJ}
    =
    -|e|L_0T
    \frac{\partial \ln G(\omega)}{\partial \omega}\Big|_{\omega=\mu_0}
    ,
    \end{equation}
which can be checked by inserting
$G(\omega)=(e^2 \pi K^2/\hbar)\sum_\sigma   \rho_\sigma^L(\omega)\rho_\sigma^R(\omega)$ into Eq.~(\ref{Eq:S_Mott}),
with $\rho_\sigma^{L/R}(\omega)$ given by Eqs.~(\ref{Eq:rho_L_approx})-(\ref{Eq:rho_R_approx}). Next, with the
use of Eq.~(\ref{Eq:L_MTJ}), one finds  the spin thermopower
    \begin{align}\label{Eq:S_MTJ}
    \mathcal{S}_\text{S}^\text{MTJ}
    =
    P_L
    \mathcal{S}^\text{MTJ}
    ,
    \end{align}
with $\mathcal{S}^\text{MTJ}$ given by Eq.~(\ref{Eq:S_final}), and the figures of merit
    \begin{equation}
    \text{ZT}^\text{MTJ}
    =
    \frac{
    \dfrac{\pi^2}{3}
    x_R^2
    T^2
    }{
    1-\dfrac{\pi^2}{3}x_R^2T^2
    }
    \ \
    \text{and}
    \ \
    \text{ZT}_\text{S}^\text{MTJ}
    =
    P_L^2
    \,
    \text{ZT}^\text{MTJ}
    .
    \end{equation}
\begin{figure}[t]
   \includegraphics[scale=0.9]{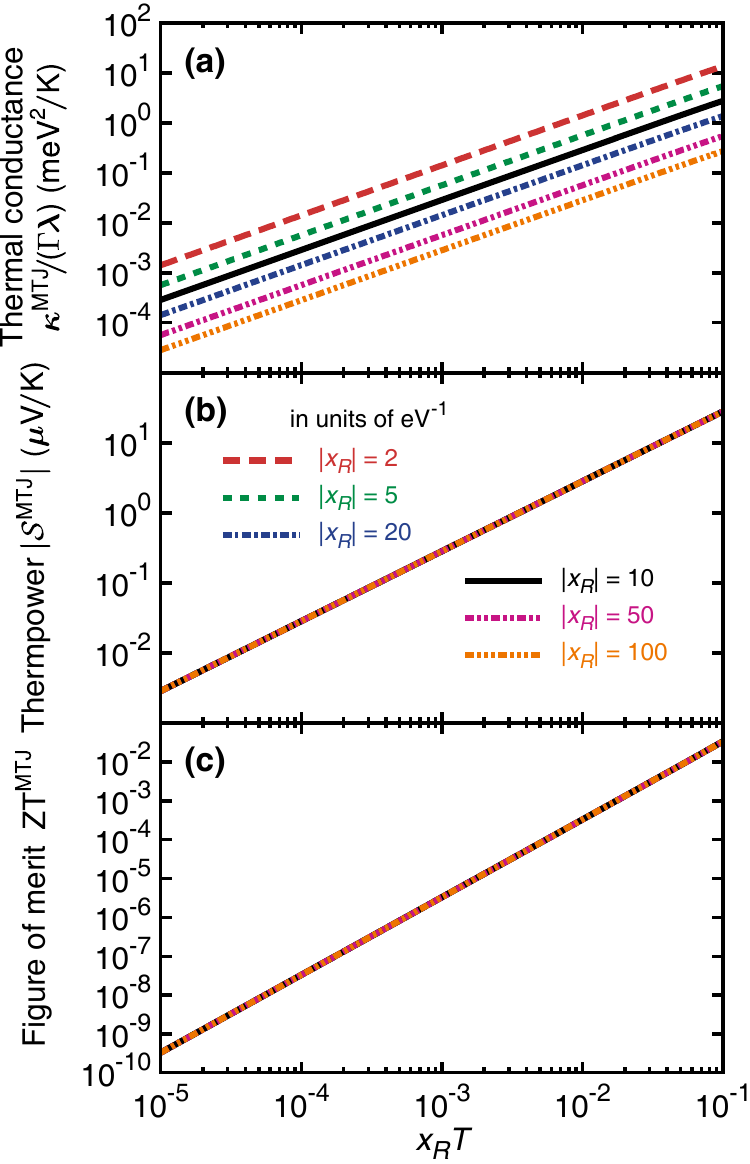}
    \caption{
    (color online) Dependence of the thermal conductance $\kappa^\text{MTJ}$,  thermopower $\mathcal{S}^\text{MTJ}$ and
    figure of merit ZT$^\text{MTJ}$ on $x_RT$ (effectively on temperature) for a junction without a magnetic impurity, shown for
    several values of the parameter $x_R$ which describes the linear term in the Taylor expansion of the right
    electrode's DOS around the Fermi level. The spin thermopower $\mathcal{S}_S^\text{MTJ}$ is now proportional to $\mathcal{S}^\text{MTJ}$.
    }
    \label{Fig:2}
\end{figure}

As one can easily note, the parameter playing a major role in the formulae above is $x_R$ describing the linear term
in the Taylor expansion of DOS in the right electrode, Eq.~(\ref{Eq:rho_R_approx}). Employing the expression
for a thermopower $\mathcal{S}$, Eq.~(\ref{Eq:S_final}), the order of magnitude for $x_R$ can be deduced from available experimental
works on thermoelectric transport in magnetic tunnel junctions and molecular junctions. Recent experiments on
the  MgO-~\cite{Walter_NatureMater.10/2011,Liebing_Phys.Rev.Lett.107/2011,Zhang_Phys.Rev.Lett.109/2012,Teixeira_Appl.Phys.Lett.102/2013,Boehnke_Rev.Sci.Instrum.84/2013} and Al$_2$O$_3$-based\cite{LeBreton_Nature475/2011,Lin_NatureCommun.3/2012} junctions show that at room temperature $|\mathcal{S}|$ can vary between a few tens of $\mu$V/K and several mV/K~\cite{Lin_NatureCommun.3/2012,Teixeira_Appl.Phys.Lett.102/2013}, with typical values oscillating around 50-200 $\mu$V/K. These agree with theoretical values found
from analytical considerations for magnon-assisted tunneling~\cite{McCann_Phys.Rev.B66/2002} and these obtained
from \emph{ab initio} studies.~\cite{Czerner_Phys.Rev.B83/2011} On the other hand, in molecular junctions
with a single fullerene molecule (i.e. C$_{60}$, PCBM or C$_{70}$)~\cite{Yee_NanoLett.11/2011,Evangeli_NanoLett.13/2013}
or  an aromatic molecule~\cite{Reddy_Science315/2007,Baheti_NanoLett.8/2008,Malen_NanoLett.9/2009_1164,Malen_NanoLett.9/2009_3406,Tan_App.Phys.Lett.96/2010,Widawsky_NanoLett.12/2011}
embedded, $\mathcal{S}$ has been observed not to exceed usually 30 $\mu$V/K at room temperature. Interestingly enough,
theoretical predictions\cite{Bergfield_NanoLett.9/2009,Bergfield_Phys.Rev.B79/2009} for some molecules from the latter
group suggest that by tuning a chemical potential  one can reach $|\mathcal{S}|$ as large as 150 $\mu$V/K. Consequently,
assuming a typical value of $|\mathcal{S}|$  at room temperature to be of the order of 100 $\mu$V/K, we find  $|x_R|\sim10$ eV$^{-1}$.

In Fig.~\ref{Fig:2} we present the thermal conductance $\kappa^\text{MTJ}$,  thermopower $\mathcal{S}^\text{MTJ}$
and figure of merit ZT$^\text{MTJ}$ as a function of $x_RT$ (effectively as a function of temperature)  for
several values of the parameter $x_R$. From the corresponding formulas follows that the
dependence on $x_RT$ is roughly linear in the temperature range where the description is valid, $x_RT\ll 1$.
Moreover, this dependence is independent on $x_R$ for thermopower and figure of merit, as follows from the
corresponding analytical formula, and is also clearly seen in Fig.~\ref{Fig:2}(b,c).
The situation is different for the thermal conductance, where different values of the parameter $x_R$ correspond to different curves,
as can be clearly seen in Fig.~\ref{Fig:2}(a). This is due to the prefactor $T$  in Eq.(\ref{Eq:kappa_MTJ}). Note, the spin thermopower
$\mathcal{S}_S^\text{MTJ}$ is now proportional to $\mathcal{S}^\text{MTJ}$, see Eq.(\ref{Eq:S_MTJ}).

\subsection{Magnetic impurity with spin\ $S$}

Let us  now turn to the situation with the spin impurity in the barrier.
Transport properties of the system depend then on a number of parameters,
including the asymmetry of the coupling  between  the impurity and electrodes, quantified in the following by $\nu_\text{as}\equiv\nu_R/\nu_L$
and the ratio $\lambda=\rho^R/\rho^L$ of the electrodes' DOS at the Fermi level. In order to facilitate
the discussion, we introduce an additional auxiliary parameter $\mathcal{A}$, defined as
    \begin{equation}
    \mathcal{A}
    =
    \lambda\nu_\text{as}^2
    ,
    \end{equation}
which describes an \emph{effective asymmetry} of the
junction containing a spin impurity.

The conductance matrix $\bm{G}$, see Eq.~(\ref{Eq:G_S_def}), can be formally separated into two parts as $\bm{G}=\bm{G}_\text{sc}+\bm{G}_\text{sf}$. The first term expressed in terms of the asymmetry parameters has the form
    \begin{align}\label{Eq:G_el}
    \renewcommand{\arraystretch}{1.75}
    \hspace*{-5pt}
    \bm{G}_\text{sc}
    &
    =
    \mathcal{T}_\text{sc}
    \bm{G}^\text{MTJ}
                \nonumber\\
    &
    =
    \Gamma
    \Bigg\{
    \!
    \lambda
    +
    \frac{\vartheta_1}{2}
    \alpha_\text{ex}^2
    \nu_L^4
    \mathcal{A}
    \sum_{\chi}
    \!
    \big(
    \mathbb{S}_{\chi\chi}^z
    \big)^{\!2}
    \!
    \Bigg\}
    \!\!
            \begin{pmatrix}
        e^2
        &
        e^2 P_L
                    \\
        \dfrac{e\hbar}{2}P_L
        &
        \dfrac{e\hbar}{2}
        \end{pmatrix}
    \hspace*{-3pt}
   \end{align}
and represents the contribution from spin-conserving electron tunneling processes, with $\bm{G}^\text{MTJ}$ given by Eq.~(\ref{Eq:G_MTJ}). The second term, in turn, takes the form
    \begin{equation}
    \renewcommand{\arraystretch}{1.75}
    \bm{G}_\text{sf}
    =
    \Gamma
    \alpha_\text{ex}^2
    \nu_L^4
        \begin{pmatrix}
        e^2\widetilde{\mathcal{T}}_\text{sf}^{(0)}
        &
        e^2 P_L\widetilde{\mathcal{T}}_\text{sf}^{(s)}
                    \\
        \dfrac{e\hbar}{2}P_L\widetilde{\mathcal{T}}_\text{sf}^{(s)}
        &
        \dfrac{e\hbar}{2}\widetilde{\mathcal{T}}_\text{sf}^{(1)}
        \end{pmatrix}
   \end{equation}
and is the contribution to the conductance which stems from spin-flip scattering of electrons on the spin impurity, with
    \begin{align}\label{Eq:tildeT0}
    \widetilde{\mathcal{T}}_\text{sf}^{(0)}
    =
    \frac{\vartheta_3}{2}
    \mathcal{A}
    -
    \vartheta_2
    \frac{
    \mathcal{A}^2
    P_L^2
    }{
    \mathcal{A}
    \big[
    \mathcal{A}+2
    \big]
    +
    \big[
    1-P_L^2
    \big]
    }
    ,
    \end{align}
    \begin{align}
    \widetilde{\mathcal{T}}_\text{sf}^{(1)}
    =\ &
    \frac{\vartheta_3}{2}
    \mathcal{A}
    \frac{
    \mathcal{A}^2
    +
    \big[
    1-P_L^2
    \big]
    }{
    \mathcal{A}
    \big[
    \mathcal{A}+2
    \big]
    +
    \big[
    1-P_L^2
    \big]
    }
                    \nonumber\\
    +\ &
    \frac{1}{4}
    \big[\vartheta_3-\vartheta_2\big]
    \frac{
    \mathcal{A}^4
    +
    \big[
    1-P_L^2
    \big]^2
    }{
    \mathcal{A}
    \big[
    \mathcal{A}+2
    \big]
    +
    \big[
    1-P_L^2
    \big]
    }
                    \nonumber\\
    +\ &
    \frac{1}{2}
    \big[\vartheta_3+\vartheta_2\big]
    \frac{
    \mathcal{A}^2
    \big[
    1-P_L^2
    \big]
    }{
    \mathcal{A}
    \big[
    \mathcal{A}+2
    \big]
    +
    \big[
    1-P_L^2
    \big]
    }
    ,
    \end{align}
    \begin{align}\label{Eq:tildeTs}
    \widetilde{\mathcal{T}}_\text{sf}^{(s)}
    =
    \frac{\vartheta_2}{2}
    \mathcal{A}
    \frac{
    \mathcal{A}^2
    -
    \big[
    1-P_L^2
    \big]
    }{
    \mathcal{A}
    \big[
    \mathcal{A}+2
    \big]
    +
    \big[
    1-P_L^2
    \big]
    }
    .
    \end{align}

In general, the processes of spin-flip scattering on the impurity correspond to opening new channels for transport through
the junction.
In the case of the spin-conserving scattering processes all the $2S+1$ channels become available
for an \emph{isotropic} spin impurity, and
each impurity state $\ket{\chi}$ gives a positive contribution to the conductance,
    $(\vartheta_1/2)
    \big(
    \alpha_\text{ex}^2\nu_L^4\mathcal{A}
    \big)
    \big(
    \mathbb{S}_{\chi\chi}^z
    \big)^{\!2}
    \geqslant
    0
    $.
For an \emph{anisotropic} impurity, in turn, only two channels  contribute to the conductance (recall the assumption discussed above).
The situation is more complicated for spin-flip scattering processes, see Eqs.~(\ref{Eq:tildeT0})-(\ref{Eq:tildeTs}), which lead to mixing of spin channels.

The thermal conductance, in turn,  is given by the following formula:
    \begin{align}\label{Eq:kappa_over_kappaMTJ}
     \frac{\kappa}{\kappa^\text{MTJ}}
    =
    1
    +
    \alpha_\text{ex}^2\nu_L^4
    \nu_\text{as}^2
    \Bigg\{
    \frac{\vartheta_1}{2}
    \sum_{\chi}
    \!
    \big(
    \mathbb{S}_{\chi\chi}^z
    \big)^{\!2}
    +
    \frac{\vartheta_3}{2}
    \Bigg\}
    \!
    .
    \end{align}
Finally, the thermopower $\mathcal{S}$ is given by Eq.~(\ref{Eq:S_final}), while the figure of merit can be written as
    \begin{multline}
    \frac{
    \text{ZT}
    }{
    \text{ZT}^\text{MTJ}
    }
    =
    1
     -
    \frac{
    2
    \alpha_\text{ex}^2\nu_L^4
    }{
    2\lambda\mathcal{T}_\text{sc}
    +
    \vartheta_3
    \alpha_\text{ex}^2\nu_L^4\mathcal{A}
    }
            \\
    \times
    \frac{
    \vartheta_2
    \mathcal{A}^2
    P_L^2
    }{
    \mathcal{A}
    \big[
    \mathcal{A}
    +
    2
    \big]
    +
    \big[
    1-P_L^2
    \big]
    }
    .
    \end{multline}
The spin thermopower $\mathcal{S}_\text{S}$, Eq.~(\ref{Eq:S_final}),  is determined by $\mathcal{S}$ and
the ratio $G_\text{S}^\text{m}/G_\text{S}$, while the spin figure of merit $\text{ZT}_\text{S}$, Eq.~(\ref{Eq:ZT_S_final}), depends on $\text{ZT}$ and the ratio
$\big(G_\text{S}^\text{m}\big)^{\!2}\!/(GG_\text{S})$. Both, $\mathcal{S}_\text{S}$ and $\text{ZT}_{\text{S}}$ can be expressed in terms of the asymmetry parameter
$\mathcal{A}$, but the corresponding formula are cumbersome and will not be presented here.

In the following we distinguish between the case of \emph{isotropic}  ($D=E=0$) and  \emph{anisotropic}  ($D\ne 0$ and $E\ne 0$) spin impurity, and we begin the discussion with the former case.

\subsubsection{The isotropic case ($D=E=0$)}

For an isotropic spin impurity, $D=E=0$, the above formulae can be further simplified.
Taking into account the explicit form of the parameters $\vartheta_n$ (see Tab.~\ref{Tab:Theta_coeff}),
one obtains
    \begin{align}\label{Eq:T_01_iso}
    \hspace*{-6pt}
    \widetilde{\mathcal{T}}_\text{sf}^{(0)}
    =
    \widetilde{\mathcal{T}}_\text{sf}^{(1)}
    =
    \frac{2}{3}S(S+1)
    \mathcal{A}
    \frac{
    \mathcal{A}^2
    +
    \big[
    1+2\mathcal{A}
    \big]
    \big[
    1-P_L^2
    \big]
    }{
    \mathcal{A}
    \big[
    \mathcal{A}+2
    \big]
    +
    \big[
    1-P_L^2
    \big]
    }
    ,
    \hspace*{-3pt}
    \end{align}
    \begin{align}\label{Eq:T_s_iso}
    \widetilde{\mathcal{T}}_\text{sf}^{(s)}
    =
    \frac{2}{3}S(S+1)
    \mathcal{A}
    \frac{
    \mathcal{A}^2
    -
    \big[
    1-P_L^2
    \big]
    }{
    \mathcal{A}
    \big[
    \mathcal{A}+2
    \big]
    +
    \big[
    1-P_L^2
    \big]
    }
    .
    \end{align}
Let us consider first the situation of a fully \emph{symmetric} junction, i.e. when $\lambda=1$ (the symmetry with respect to DOS, $\rho^L=\rho^R$) and $\nu_\text{as}=1$ (the symmetry with respect to the coupling between the impurity and electrodes, $\nu_L=\nu_R=1$), which corresponds to  $\mathcal{A}=1$.
The above formulae reduce then to the following ones:
    \begin{equation}
    \widetilde{\mathcal{T}}_\text{sf}^{(0)}
    =
    \widetilde{\mathcal{T}}_\text{sf}^{(1)}
    =
    \frac{2}{3}S(S+1)
    \frac{4-3P_L^2}{4-P_L^2}
    ,
    \end{equation}
     \begin{equation}
     \widetilde{\mathcal{T}}_\text{sf}^{(s)}
     =
     \frac{2}{3}S(S+1)
     \frac{P_L^2}{4-P_L^2}
     ,
     \end{equation}
so that the conductance matrix takes the form
     \begin{align}\label{Eq:G_A_1}
     \hspace*{-5pt}
     \bm{G}
     &
     = \Gamma
     \Bigg\{
     \!
     1
     +
     \frac{1}{3}
     S(S+1)
     \alpha_\text{ex}^2
     \Bigg\}
     \!\!
     \renewcommand{\arraystretch}{1.5}
         \begin{pmatrix}
         e^2
         &
         e^2
         P_L
                     \\
         \dfrac{e\hbar}{2}
         P_L
         &
         \dfrac{e\hbar}{2}
         \end{pmatrix}
                     \nonumber
     \\
     &+
     \frac{2}{3}
     \Gamma
     S(S+1)
     \alpha_\text{ex}^2
      \renewcommand{\arraystretch}{2.5}
         \begin{pmatrix}
         e^2\dfrac{4-3P_L^2}{4-P_L^2}
         &
         e^2 \dfrac{P_L^2}{4-P_L^2}
                     \\
         \dfrac{e\hbar}{2} \dfrac{P_L^2}{4-P_L^2}
         &
         \dfrac{e\hbar}{2}\dfrac{4-3P_L^2}{4-P_L^2}
         \end{pmatrix}
     \!\!
     .
     \hspace*{-2pt}
    \end{align}
Here,  the first and second terms represent the spin-conserving and spin-flip parts of
the conductance, respectively.
In turn, the heat conductance can be written as
    \begin{align}
     \frac{\kappa}{\kappa^\text{MTJ}}
    =
    1
    +
    S(S+1)
    \alpha_\text{ex}^2
    .
    \end{align}
The  formulae above have been used to get some numerical results to be discussed below.
In the case of spin thermopower and spin figure of merit, on the other hand, to obtain the numerical results we used the general expressions~(\ref{Eq:S_final})-(\ref{Eq:ZT_S_final}).

\begin{figure}[t]
   \includegraphics[width=0.99\columnwidth]{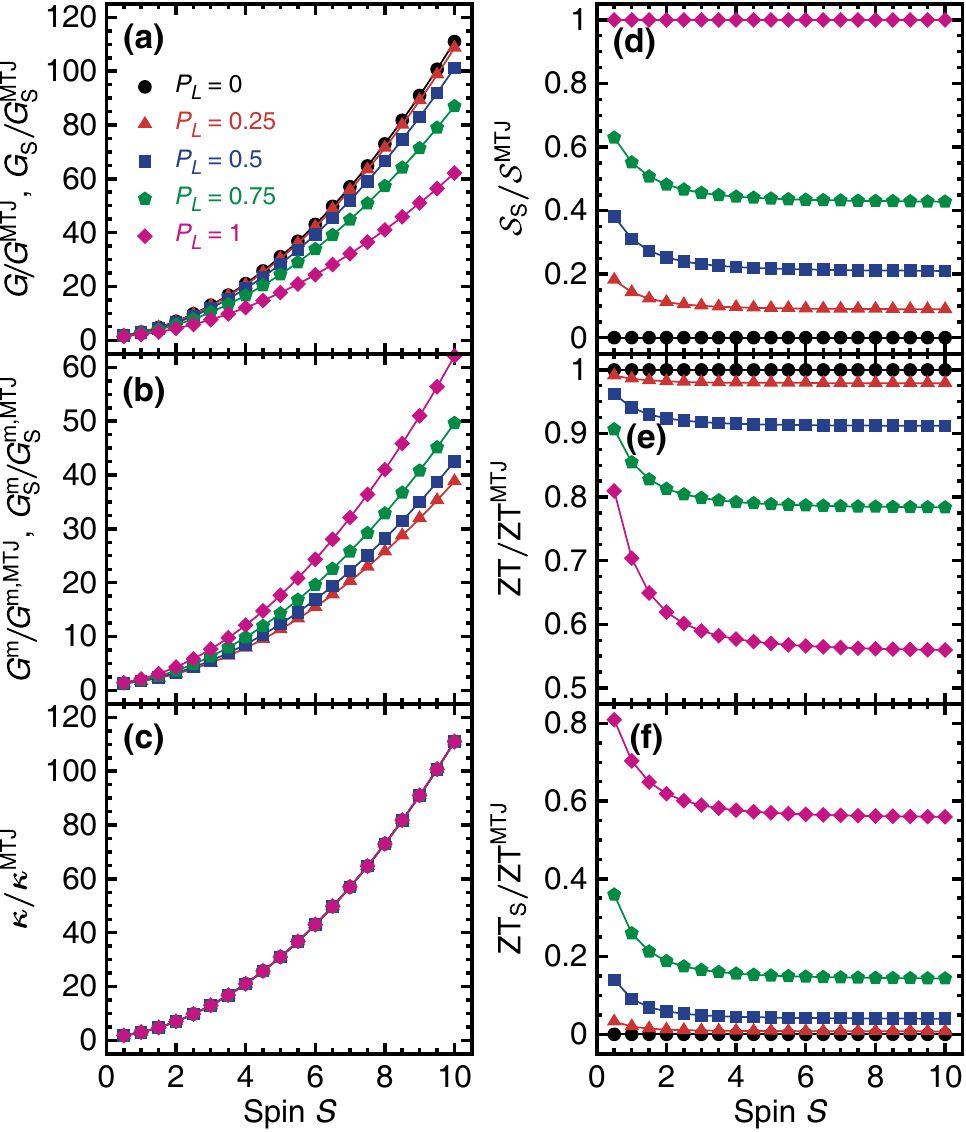}
    \caption{
    (color online) Diagonal (a) and nondiagonal (b) elements of the conduction matrix $\bm{G}$, electronic contribution to the heat conductance $\kappa$ (c),
     spin thermopower $\mathcal{S}_\text{S}$ (d), figure of merit $\text{ZT}$ (e), and spin figure of merit $\text{ZT}_\text{S}$ (f) shown as functions of the
     impurity spin number $S$ for several values of the left electrode's spin-polarization parameter $P_L$.
     Note that the points correspond to spin numbers while the lines serve merely as a guide for eyes.
     Remaining parameters: $T=1$ K, $x_R=10$ eV$^{-1}$, $\alpha_\text{d}=\alpha_\text{ex}=1$, $\lambda=1$ and $\nu_L=\nu_R=1$,
      so that $\mathcal{A}=1$ (i.e. the junction is fully \emph{symmetric} with respect to DOS and exchange coupling).
        }
    \label{Fig:3}
\end{figure}

\begin{figure}[t]
   \includegraphics[width=0.99\columnwidth]{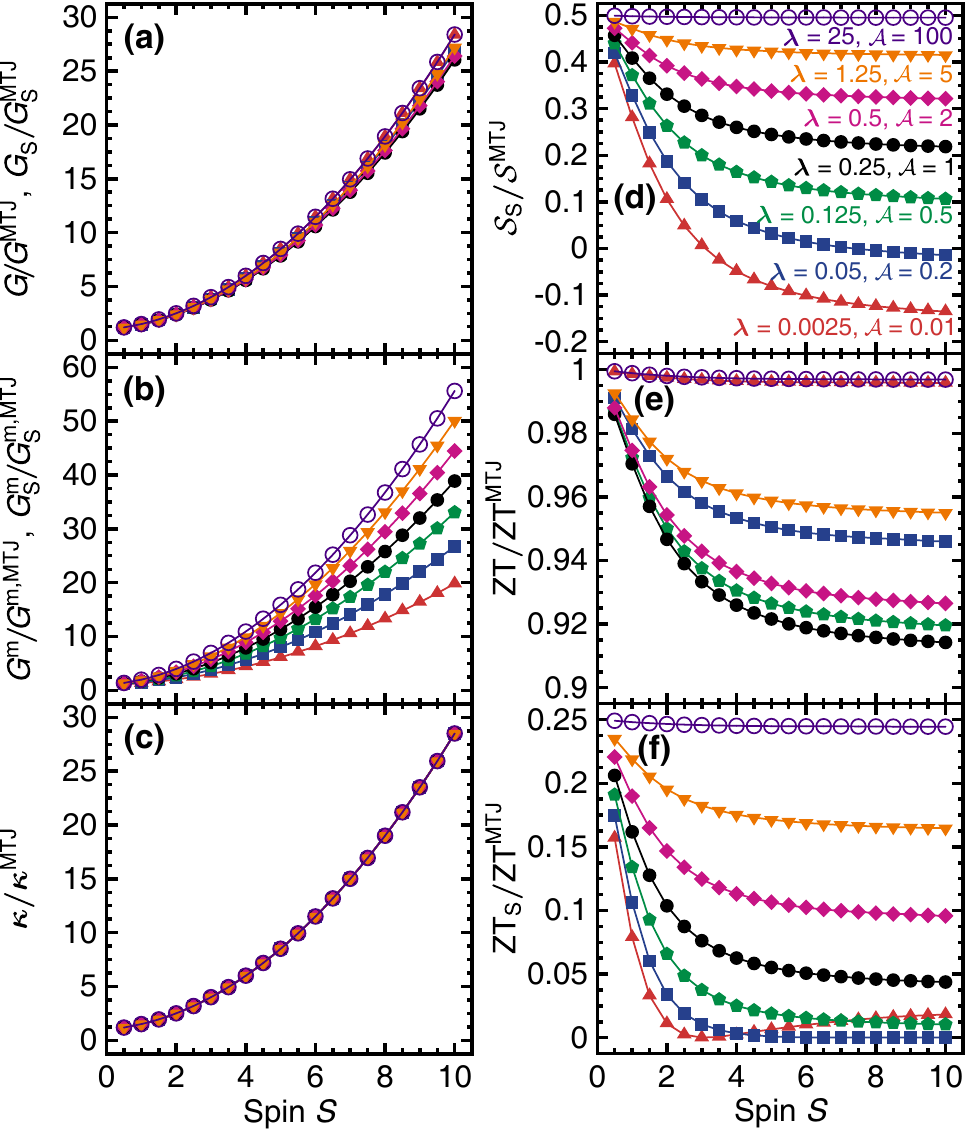}
    \caption{
    (color online) Analogous to Fig.~\ref{Fig:3}, but now the quantities under consideration are
    presented for different values of the parameter $\lambda$ describing the asymmetry of DOS at the Fermi level on
    opposite sides of the junction. In addition, we assume that the spin impurity is more coupled to the right electrode,
    i.e. $2\nu_L=\nu_R=1$, so that the effective asymmetry parameter $\mathcal{A}$ is related to $\lambda$ as
    $\mathcal{A}=4\lambda$. Except $P_L=0.5$, all other parameters as in Fig.~\ref{Fig:3}.
        }
    \label{Fig:4}
\end{figure}

The corresponding results
 for a \emph{symmetric} junction,  $\mathcal{A}=1$ with $\lambda=\nu_\text{as}=1$,
 are shown in Fig.~\ref{Fig:3}, where  all the elements of the conductance matrix $\bm{G}$, electronic contribution
to the heat conductance $\kappa$, spin thermopower $\mathcal{S}_\text{S}$, and figures
of merit $\text{ZT}$ and $\text{ZT}_{\text{S}}$, normalized to the corresponding quantities for MTJ without impurity,
 are shown as a function of the spin number $S$ for indicated values of the polarization factor $P_L$.
Note that the electrical conductance $G$ is
proportional to the spin conductance $G_\text{S}$, and the  nondiagonal conductance $G^\text{m}$ is proportional to $G^\text{m}_\text{S}$.
For the lowest value of $S$, i.e. $S=1/2$, the contributions from direct tunneling and exchange terms are comparable
and the total conductances are rather small.
The role of exchange term in the tunneling Hamiltonian increases with increasing $S$
and the conductances grow  as $\sim S^2$ with increasing $S$. This behavior is clear as the tunneling probability with exchange interaction between
the electron and impurity is effectively proportional to $S^2$, see the tunneling Hamiltonian given by Eq.(\ref{Eq:Ham_T}).
Worth of note is that the nondiagonal conductances ($G^\text{m}$ and $G_\text{S}^\text{m}$) achieve their maximal values for $P_L=1$,
while the diagonal ones ($G$ and $G_\text{S}$) for $P_L=0$, as one might expect.
Additionally, if the left electrode is nonmagnetic ($P_L=0$) the nondiagonal components of the conductance matrix $\bm{G}$ vanish.

Unlike the electric conductance $G$,
the electronic contribution to the heat conductance $\kappa$, shown in Fig.~\ref{Fig:3}(c), is independent of the polarization $P_L$, though it
grows with $S$ similarly as $G$ does.
By contrast,
the spin thermopower  and both figures of merit  are decreased in comparison to the corresponding values for MTJ
without the impurity.  However, the reduction in the values of $\mathcal{S}_\text{S}$ and $\text{ZT}_\text{S}$ is  inversely proportional
to the spin-polarization of the left electrode, i.e.  for $P_L\rightarrow0$ one obtains $\mathcal{S}_\text{S}^{\phantom{X}}\!\!/\mathcal{S}_\text{S}^\text{MTJ}\rightarrow0$
and $\text{ZT}_\text{S}^{\phantom{X}}\!\!/\text{ZT}_\text{S}^\text{MTJ}\rightarrow0$, whereas the conventional (charge) figure of merit $\text{ZT}$ is generally
diminished as $P_L$ grows.
Interestingly, the spin thermopower becomes reduced with increasing spin number $S$, which appears due to enhanced spin
mixing by spin-flip transmission processes. The corresponding spin figure of merit becomes also suppressed with increasing spin number.

\begin{figure}[t]
   \includegraphics[width=0.95\columnwidth]{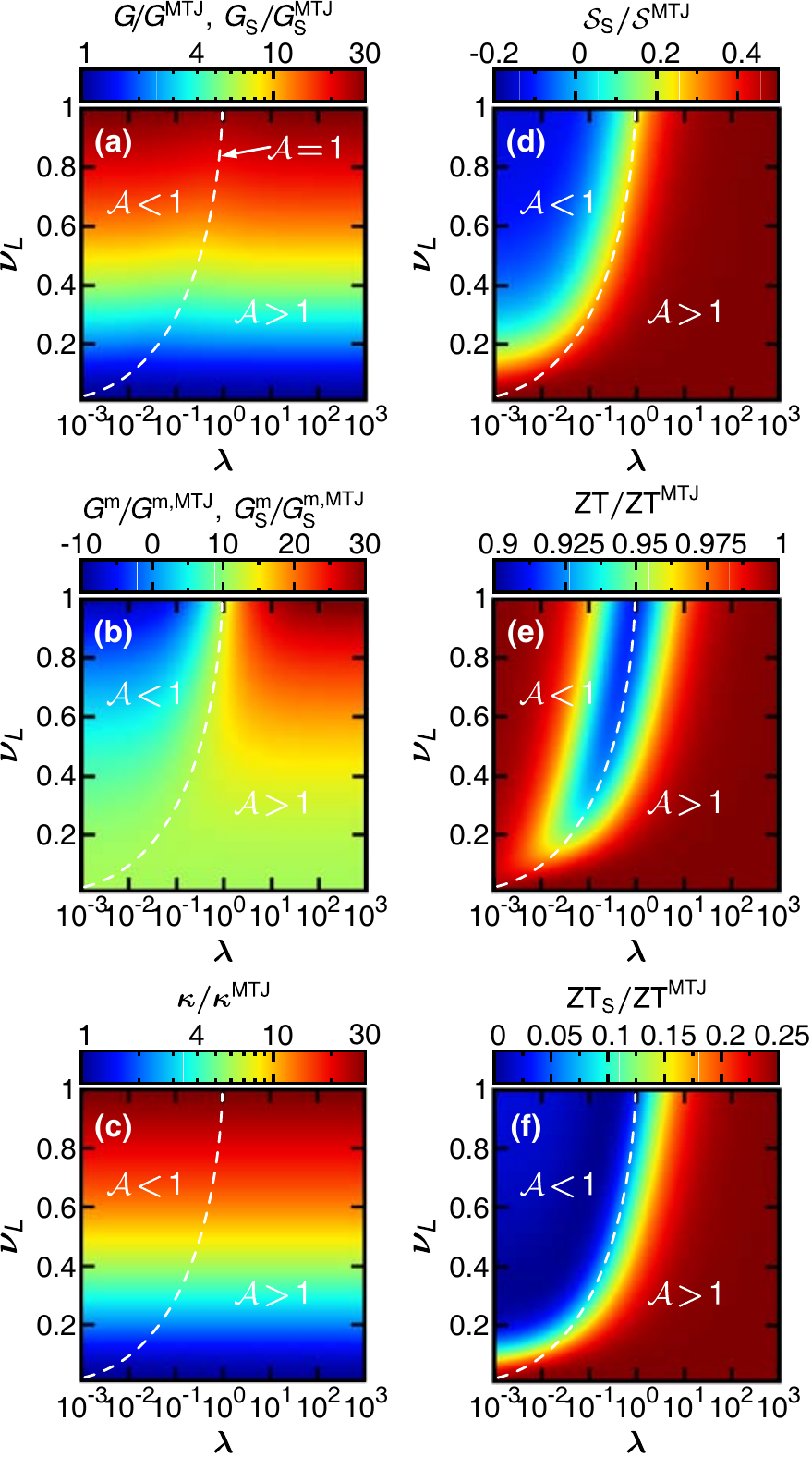}
    \caption{
    (color online) Analogous to Fig.~\ref{Fig:4}, but now the quantities under consideration are plotted as a
    function of the effective asymmetry parameter $\mathcal{A}$.
     This is achieved by varying parameters $\lambda$  and $\nu_\text{as}$ (to be precise, we change $\nu_L$ while keeping $\nu_R=1$).
     The white dashed line represents  $\mathcal{A}=1$.
     Note that the case of a fully \emph{symmetric} junction, i.e. for $\mathcal{A}=1$ with $\lambda=\nu_\text{as}=1$, corresponds in each plot to the point with coordinates $(\lambda=1,\nu_L=1)$.
      Except $P_L=0.5$ and $S=5$, all parameters as in Fig.~\ref{Fig:4}.
    }
    \label{Fig:5}
\end{figure}

Let us now look in more details at the case of an \emph{asymmetric} junction, $\mathcal{A}\ne 1$.
Figure~\ref{Fig:4} presents
the relative conductance matrix elements, heat conductance,
spin thermopower, and
figures of merit  as a function of spin number $S$, similarly as in Fig.~\ref{Fig:3} but for different values of the parameters
describing the junction asymmetry.
First of all, one can immediately notice that
the relative diagonal conductances, i.e. charge  ($G/G^\text{MTJ}$) and spin ($G_\text{S}^{\phantom{X}\!\!}/G^\text{MTJ}_\text{S}$) ones shown in Fig.~\ref{Fig:4}(a),
are only weakly affected by
 the change in asymmetry
parameters assumed in  Fig.~\ref{Fig:4}. This dependence is much more pronounced for relative nondiagonal elements of
the conductance matrix in Fig.~\ref{Fig:4}(b), also cf. Eqs.~(\ref{Eq:T_01_iso})-(\ref{Eq:T_s_iso}),
where a significant asymmetry of the junction leads to increase of the relevant components of $\bm{G}$.
Apart from this, the variation of all the conductances with  the spin
number $S$
resembles that observed earlier
 in  Fig.~\ref{Fig:3}, so it will not be discussed here.
On the other hand, according to Eq.~(\ref{Eq:kappa_over_kappaMTJ}) and Fig.~\ref{Fig:4}(c), the relative heat conductance
is independent of the asymmetry parameter $\mathcal{A}$. However,  one should bear in mind that $\kappa^\text{MTJ}$ as well as elements of the conductance matrix $\bm{G}^\text{MTJ}$ are actually sensitive to the asymmetry of the electrodes' DOS at the Fermi level characterized by $\lambda$, see Eq.~(\ref{Eq:kappa_MTJ})
and Eq.~(\ref{Eq:G_MTJ}).

The relative spin thermopower, shown in Fig.~\ref{Fig:4}(d),  depends remarkably on the asymmetry parameters:
whereas in the limit of  \emph{large} $\mathcal{A}$ [i.e. $\mathcal{A}\gg 1$, see open circles in Fig.~\ref{Fig:4}(d)] $\mathcal{S}_\text{S}^{\phantom{X}}\!\!/\mathcal{S}^\text{MTJ}_\text{S}\approx \text{const.}$ regardless of the value of spin number~$S$,
for \emph{small} $\mathcal{A}$ [i.e. $0<\mathcal{A}\ll 1$, triangles in Fig.~\ref{Fig:4}(d)] $\mathcal{S}_\text{S}$ significantly varies with $S$ and it can even change its sign.
Employing Eq.~(\ref{Eq:S_final}) one derives
    \begin{equation}\label{Eq:S_S_asymmetry_D_0}
    \hspace*{-1pt}
    \frac{\mathcal{S}_\text{S}}{\mathcal{S}^\text{MTJ}_S}
    =
    \frac{G_\text{S}^\text{m}}{G_\text{S}}
    =
    P_L
    \frac{
    \mathcal{X}_\text{sc}
    +
    \dfrac{
    \mathcal{A}^2
    -
    \big[
    1-P_L^2
    \big]
    }{
    \mathcal{A}
    \big[
    \mathcal{A}+2
    \big]
    +
    \big[
    1-P_L^2
    \big]
    }
    }{
    \mathcal{X}_\text{sc}
    +
    \dfrac{
    \mathcal{A}^2
    +
    \big[
    1+2\mathcal{A}
    \big]
    \big[
    1-P_L^2
    \big]
    }{
    \mathcal{A}
    \big[
    \mathcal{A}+2
    \big]
    +
    \big[
    1-P_L^2
    \big]
    }
    }
    ,
    \end{equation}
where the coefficient $\mathcal{X}_\text{sc}$
is given by
    \begin{equation}
    \mathcal{X}_\text{sc}
    =
    \frac{
    3\lambda+
    S(S+1)
    \alpha_\text{ex}^2\nu_L^4\mathcal{A}
    }{
    2S(S+1)
    \alpha_\text{ex}^2\nu_L^4\mathcal{A}
    }
     .
    \end{equation}
Since when plotting Fig.~\ref{Fig:4} we assumed that the parameter $\nu_\text{as}$, describing the asymmetry of coupling between the impurity and electrodes, is kept constant, it means that $\mathcal{A}$ can in fact be identified with $\lambda$.
As a result, the  disparate behavior of the spin thermopower can be understood on the basis of the inequality of $\rho^L$ and $\rho^R$.
It's clear from Eqs.~(\ref{Eq:G_el})-(\ref{Eq:tildeTs}) that whereas the spin-conserving parts of spin conductances depend on the DOSs of electrodes in a straightforward way,
$G_\text{S,sc}^\text{m}=P_LG_\text{S,sc}\propto\rho^L\rho^R$, its spin-flip counterparts exhibit a nontrivial dependence on these two parameters. Because $\mathcal{S}_\text{S}\propto G_\text{S}^\text{m}/G_\text{S}^{\phantom{X}}\!$ [note that $\mathcal{S}^\text{MTJ}$ is only proportional to $x_R(\rho^R)$], its sensitivity to variations of the ratio $\rho^R/\rho^L$ therefore originates from the spin-flip electron transport, see Eq.~(\ref{Eq:S_S_asymmetry_D_0}).
In particular, for \emph{large} $\mathcal{A}$ corresponding to $\rho^R \gg \rho^L$, one gets
    \begin{equation}
    \frac{G_\text{S,sf}^\text{m}}{G_\text{S}^\text{MTJ}}
    =
    P_L\frac{G_\text{S,sf}}{G_\text{S}^\text{MTJ}}
    =
     \frac{2}{3}
     P_L
     S(S+1)
     \alpha_\text{ex}^2\nu_L^4\nu_\text{as}^2
     ,
    \end{equation}
while for  \emph{small} $\mathcal{A}$ corresponding to $\rho^R \ll \rho^L$,
    \begin{equation}
    \frac{G_\text{S,sf}^\text{m}}{G_\text{S}^\text{MTJ}}
    =
    -P_L\frac{G_\text{S,sf}}{G_\text{S}^\text{MTJ}}
    =
    -
    \frac{2}{3}
    P_L
     S(S+1)
    \alpha_\text{ex}^2\nu_L^4\nu_\text{as}^2
     ,
    \end{equation}
so that in the limit of very large $S$ the following approximate expression for the spin thermopower is found
    \begin{equation}
    \left\{
            \begin{aligned}
            \lim\limits_{\text{large}\, \mathcal{A}}
            \mathcal{S}_\text{S}^{\phantom{X}}\!\!/\mathcal{S}^\text{MTJ}_\text{S}
            &=
            P_L,
            \\
            \lim\limits_{\text{small}\, \mathcal{A}}
            \mathcal{S}_\text{S}^{\phantom{X}}\!\!/\mathcal{S}^\text{MTJ}_\text{S}
            &=
            -\frac{1}{3}P_L
            .
            \end{aligned}
    \right.
    \end{equation}
Interestingly, one finds $\mathcal{S}_\text{S}=0$ when $G_\text{S}^\text{m}=0$,
which  means that the
vanishing of the spin thermopower in Fig.~\ref{Fig:4}(d) occurs for parameters
precluding the flow of spin current when an electric bias is applied, i.e. the corresponding electric current is not spin-polarized.

In order to complete the above discussion of how the asymmetry of the junction with a magnetic impurity affects its thermoelectric properties,
we show in Fig.~\ref{Fig:5} the dependence of relevant thermoelectric coefficients on both the asymmetry parameters $\lambda$ and $\nu_\text{as}$.
The white dashed lines signify  there the case of $\mathcal{A}=1$, separating the regions of \emph{large} $\mathcal{A}$ ($\mathcal{A}>1$) and  \emph{small} $\mathcal{A}$ ($\mathcal{A}<1$).
One can note  that the conventional figure of merit $\text{ZT}$ departs from the corresponding value  $\text{ZT}^\text{MTJ}$  for an empty junction only when
$\mathcal{A}$ only slightly differs from 1.

\subsubsection{The anisotropic case ($D\neq0$ and $E\neq0$)}

\begin{figure}[t]
   \includegraphics[width=0.975\columnwidth]{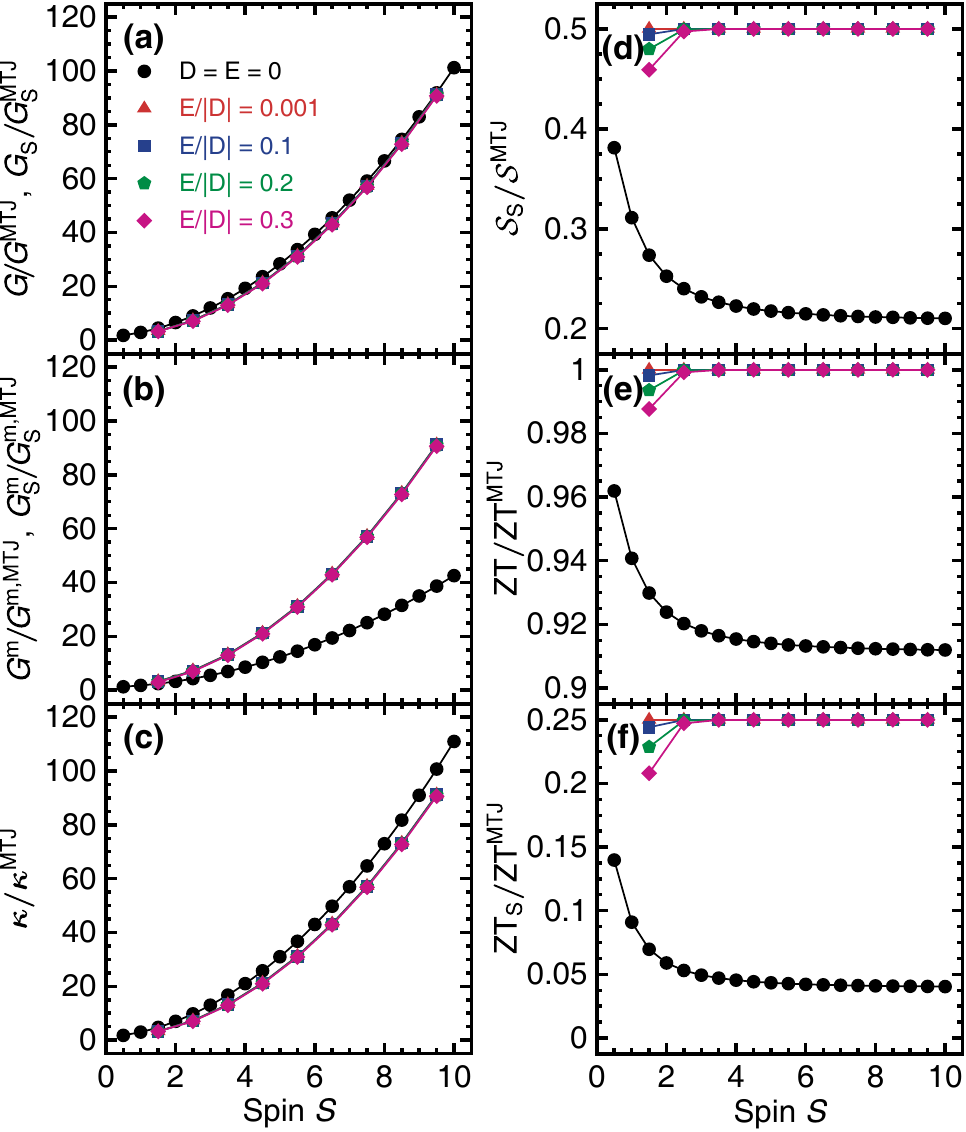}
    \caption{
    (color online)   Analogous to Fig.~\ref{Fig:3}, but now the quantities under consideration are plotted for the
    case of an \emph{anisotropic} spin impurity with $D=100$ $\mu$eV and indicated values of the transverse magnetic anisotropy
    $E$.  Except $P_L=0.5$ and $T=0.1$ K, all parameters as in Fig.~\ref{Fig:3}. For comparison, black dots representing the situation of an \emph{isotropic}
    ($D=E=0$) spin impurity are also presented.
    }
        \label{Fig:6}
     \end{figure}

Let's  now consider the general case of anisotropic magnetic impurity, $D\ne0$ and $E\ne 0$. Owing to the transverse anisotropy,
electrons tunneling through the barrier can reverse its spin orientation in the linear response and low temperature regimes,
as we have already mentioned above. We recall  that
for the anisotropic case we assume that only the  states of the ground dublet $\ket{\chi_{\pm S}}$ participate in transport, which basically means that
results of this section apply for $T \ll (2S-1)D$.
The corresponding numerical results for the conductance matrix elements, heat conductance, spin thermopower and both figures of merit
are presented in Fig.~\ref{Fig:6} for
a representative value of the
 uniaxial anisotropy constant $D$ and several indicated values of the transverse anisotropy $E$. For comparison, we also show there the
corresponding results for the case
of $D=E=0$.
Since the matrix elements $\big|\mathbb{S}_{\chi_{-S}\chi_S}^\pm\big|$ and accordingly the
coefficients $\vartheta_n$ for $n=2,3$ (see Tab.~\ref{Tab:Theta_coeff}) are usually small,
the diagonal and nondiagonal conductances as well as the heat conductance are almost independent on the transverse anisotropy. However, they differ
from the corresponding parameters in the limit of isotropic impurity. This difference is relatively small in the case of diagonal conductance matrix elements
and the heat conductance, but becomes remarkable for nondiagonal conductance elements.
This may be accounted for by taking into account that though the summation is now over two states of lowest energy, the corresponding occupation probability
for the two states is accordingly enhanced as the excited states are not occupied for $T \ll (2S-1)D$.

Similarly, the spin thermopower and figures of merit are
also weakly dependent on the transverse anisotropy, but significantly differ from the corresponding parameters in the  isotropic limit.
The spin thermopower and both figures of merit are enhanced in comparison to those in the limit of
$D=E=0$. Interestingly, the dependence of spin thermopower and figures of merit on the spin number $S$ in the anisotropic case is different
from that in the isotropic one. These parameters initially increase with increasing $S$ for small values of $S$, and then become
independent on $S$ with a further increase in $S$. This difference follows from the fact that only two states of lowest energy are included
in the anisotropic case due to the energy barrier, while all spin states of the impurity are taken into account
in the isotropic case as they all are degenerate.

\subsection{Limit of no charge transport between different electrodes}

Analysis of Eqs.~(\ref{Eq:L_final})-(\ref{Eq:T_s_final}) leads to the interesting conclusion that even if no transport of electrons is allowed
between two different electrodes, i.e. for $\alpha_\text{d}=0$ and $\alpha_\text{ex}^{LR}=0$, one can still have nonzero transport of spin, which is
stimulated by a spin bias $\delta V_\text{S}$, for the contribution to $\bm{G}_\text{sf}$ from $\mathcal{T}_\text{sf}^{(1)}$
[specifically, see the second term of Eq.~(\ref{Eq:T_1_final})]
remains nonzero,
    \begin{align}\label{Eq:G_S_noTUN}
    G_\text{S}
    =
    \frac{e\hbar}{2}
    \Gamma
    \alpha_\text{ex}^2\nu_L^4
    \Bigg\{
    &
    \frac{1}{4}
    \big[\vartheta_3-\vartheta_2\big]
    \frac{
    \mathcal{A}^4
    +
    \big[
    1-P_L^2
    \big]^2
    }{
    \mathcal{A}^2
    +
    \big[
    1-P_L^2
    \big]
    }
    \nonumber
                    \\
    +\
    &
    \frac{1}{2}
    \big[\vartheta_3+\vartheta_2\big]
    \frac{
    \mathcal{A}^2
    \big[
    1-P_L^2
    \big]
    }{
    \mathcal{A}^2
    +
    \big[
    1-P_L^2
    \big]
    }
    \Bigg\}
    .
    \end{align}
The nonzero spin current in the absence of charge current appears due to single-electrode tunneling processes.
In particular, of key importance are the processes in which an electron
scattering on the impurity spin
changes its spin and thus transfers to/from
the impurity  a quantum of angular momentum. If a spin bias is applied, such processes transfer effectively angular
momentum from one electrode to the impurity, and then from impurity to the other electrode. This, in turn, gives rise to
a resultant spin current flowing through the junction. One may say that spin angular momentum is effectively
pumped between the electrodes without actual charge transport across the junction. It is worth noting that the
crucial role in the process under discussion is played by the spin impurity which serves  as an intermediate reservoir of angular
momentum.~\cite{Ren_arXiv:1310.4222/2013}

\begin{figure}[t!]
   \includegraphics[width=0.975\columnwidth]{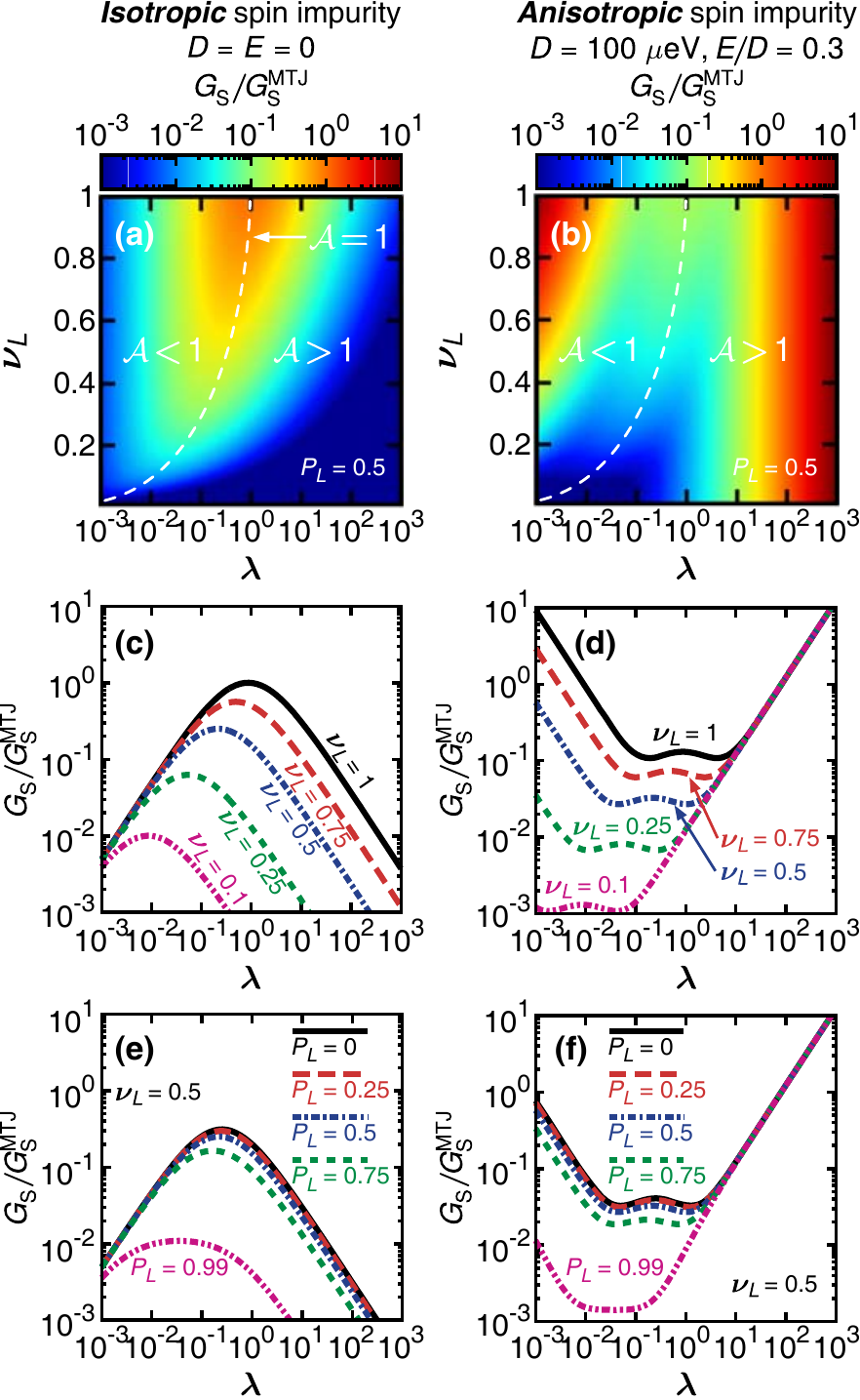}
    \caption{
    (color online) Spin transport in the absence of electron tunneling between left and right electrodes for the impurity of
    spin $S=3/2$. (a)-(b) Linear-response spin conductance $G_\text{S}$ shown as a function of the parameters $\lambda$
    (the asymmetry of right/left DOS at the Fermi level) and $\nu_L$ (the coupling of the spin impurity to the left electrode)
    for the spin polarization of the left electrode $P_L=0.5$.
    Note that for practical reasons, here we normalize $G_\text{S}$ by its value $G_\text{S}^\text{MTJ}=(e\hbar/2)\Gamma\lambda$ for the case of a junction without the impurity.
    However, one should bear in mind the crucial difference between $G_\text{S}$ and $G_\text{S}^\text{MTJ}$, namely that the latter  is associated with charge transport whereas the former is not.
    Panels (c) and (d) present cross-sections of (a) and (b), respectively,
    for indicated values of $\nu_L$. (e)-(f) Spin conductance $G_\text{S}$  plotted for different values of the spin-polarization
    parameter $P_L$ and $\nu_L=0.5$. \emph{Left/right panel} corresponds to the \emph{isotropic/anisotropic} spin impurity.
    Except $T=0.1$ K and $\nu_R=1$, all parameters as~in~Fig.~\ref{Fig:3}.
     }
    \label{Fig:7}
\end{figure}

Numerical results on the spin conductance $G_\text{S}$ in the absence of charge transport
are shown in Fig.~\ref{Fig:7} as a function of the asymmetry parameters for both
\emph{isotropic} ($D=E=0$) and \emph{anisotropic} ($D\neq0$ and $E\neq0$) situations.
For the sake of conceptual simplicity, we focus the discussion on a possibly smallest, and nontrivial from the point of view of magnetic anisotropy, half-integer value of the impurity spin $S=3/2$. Furthermore, in order to assess the efficiency of the spin-transport processes under consideration, we relate $G_\text{S}$ to
the spin conductance $G_\text{S}^\text{MTJ}=(e\hbar/2)\Gamma\lambda$ of a bare junction, i.e. without an impurity (recall Sec.~\ref{Sec:NO_impurity}). In that case, however, transport of spin between electrodes occurs entirely as a result of the charge transfer, whereas at present no tunneling of electrons across the junction takes place.

First of all, it can be notice that for the isotropic spin impurity the maximum value of $G_\text{S}^{\phantom{X}}\!\!/G_\text{S}^\text{MTJ}$ is reached for $\mathcal{A}=1$, whereas  for the anisotropic one at this point only a local maximum develops.
The origin of this  difference can be explained as follows. From the analysis of Eq.~(\ref{Eq:G_S_noTUN}) it stems that the magnetic properties of the spin impurity enter the expression exclusively \emph{via} the coefficients $\vartheta_2$ and $\vartheta_3$, see Tab.~\ref{Tab:Theta_coeff}.
Importantly, in the \emph{isotropic} case $\vartheta_2=\vartheta_3=(4/3)S(S+1)$, so that only the second term of Eq.~(\ref{Eq:G_S_noTUN}) survives, and for $S=3/2$ we obtain
$
\big[\vartheta_3+\vartheta_2\big]/2=5
$.
On the other hand, in the \emph{anisotropic} case $\vartheta_2\neq\vartheta_3$, and using definitions given in Tab.~\ref{Tab:Theta_coeff} we find
    \begin{equation}
    \left\{
        \begin{aligned}
        \frac{1}{4}\big[\vartheta_3-\vartheta_2\big]
        &=
        \big|\mathbb{S}_{\chi_{-S}\chi_S}^+\big|^2
        \big|\mathbb{S}_{\chi_{-S}\chi_S}^-\big|^2
        ,
        \\
        \frac{1}{2}\big[\vartheta_3+\vartheta_2\big]
        &=
        \big|\mathbb{S}_{\chi_{-S}\chi_S}^+\big|^4
        +
        \big|\mathbb{S}_{\chi_{-S}\chi_S}^-\big|^4
        ,
        \end{aligned}
    \right.
    \end{equation}
with
    \begin{equation}
    \big|\mathbb{S}_{\chi_{-S}\chi_S}^+\big|^2
    =
    \frac{
    \Big(D-\sqrt{D^2+3E^2}\Big)^2
    }{
    D^2+3E^2
    }
    ,
    \end{equation}
and
    \begin{equation}
    \big|\mathbb{S}_{\chi_{-S}\chi_S}^-\big|^2
    =
    \frac{
    9E^2
    }{
    D^2+3E^2
    }
    .
    \end{equation}
We emphasize that the two above formulae hold only for $S=3/2$, and no generalization to an arbitrary $S$ is possible. For
the anisotropy parameters $D$ and $E$ used in the right panel of Fig.~\ref{Fig:7} we get
$
\big[\vartheta_3-\vartheta_2\big]/4\approx0.008
$
and
$
\big[\vartheta_3+\vartheta_2\big]/2\approx0.4
$.
It can also be easily checked that  the absence of transverse magnetic anisotropy ($E=0$) leads to $\big|\mathbb{S}_{\chi_{-S}\chi_S}^\pm\big|^2=0$.
In consequence, for $\mathcal{A}\approx1$ the first term of $G_\text{S}$, Eq.~(\ref{Eq:G_S_noTUN}), contributes negligibly.
The situation changes notably when $\mathcal{A}$ differs significantly from 1 and the term in question becomes determinative.
The key observation is that whereas for the \emph{isotropic} spin impurity the maximal value of the spin conductance  $G_\text{S}^{\phantom{X}}\!\!=G_\text{S}^\text{MTJ}$ is achieved only for a symmetric junction ($\mathcal{A}=1$ for $\lambda=\nu_\text{as}=1$) and $G_\text{S}^{\phantom{X}}\!\!<G_\text{S}^\text{MTJ}$ otherwise, see Fig.~\ref{Fig:7}(a,c), for the \emph{anisotropic} spin impurity it's possible to obtain $G_\text{S}^{\phantom{X}}\!\!>G_\text{S}^\text{MTJ}$ by allowing for the asymmetry of the junction, see Fig.~\ref{Fig:7}(b,d).
As a result, the pumping of angular momentum between electrodes \emph{via} the anisotropic spin impurity without charge transport seems to be more effective that the spin transport associated with charge transport in a conventional magnetic tunnel junction.
In addition, it is interesting to note that the spin conductance decreases with
increasing spin polarization $P_L$ of the electrode, and achieves maximum for zero spin polarization of the left electrode, $P_L=0$.

\section{\label{Sec:Conclusions}Summary  and conclusions}
We have considered electronic transport and thermoelectric properties of a magnetic tunnel junction with a single magnetic impurity
embedded in the barrier.
This corresponds, for instance, to a magnetic tip above a molecule located on a substrate.
The molecule was described by a giant spin Hamiltonian with uniaxial and transverse components of magnetic anisotropy, while
the tunneling Hamiltonian was taken in the form which included direct tunneling between the electrodes as well as
tunneling with exchange interaction between the electrons and molecule.

The key objective was a description of spin related effects in electronic transport and thermoelectricity, like spin Peltier and spin
Seebeck effects. The analysis  was limited to a linear response regime and elastic scattering processes.
A particular attention was paid to
the role of spin-flip scattering of electrons on the impurity during tunneling between the electrodes. Two situations were distinguished: \emph{isotropic}, $D=E=0$, and \emph{anisotropic},
$D\ne 0$ and $E\ne 0$, ones. To exclude inelastic tunneling processes, the temperature in the anisotropic case was limited to
temperatures smaller than
 the zero-field splitting energy.

We have shown that the transverse anisotropy (for an \emph{anisotropic} spin impurity) has a minor influence on the charge and spin conductance as well as on the thermal conductance and
thermoelectric parameters (spin thermopower and spin figure of merit). However, these transport and thermoelectric coefficients differ
from the corresponding ones in the \emph{isotropic} case, and this difference is especially remarkable for the nondiagonal elements of the conductance matrix.
The difference in the conductances stems from the anisotropy barrier which appears in the anisotropic case and which limits the number of states participating in spin-conserving transport processes. In the case of an \emph{ isotropic} spin impurity, on the other hand, all states contribute to transport. It is also worth noting that the ratio of thermal conductance in the isotropic case to the thermal conductance of a bare  junction  (i.e. without a spin impurity) is independent of the electrode's polarization. The corresponding ratio for diagonal and nondiagonal elements of the conductance matrix depend on spin polarization $P_L$, and this ratio for nondiagonal elements reaches maximum for $P_L=1$, while for diagonal elements for $P_L=0$.

As a special case, we have also analyzed the situation when a spin current stimulated by a spin bias can flow through the junction in the absence of  a charge current.
This corresponds to the situation when the external electrodes are exchange-coupled to the molecule, and there are no electron
tunneling processes (neither direct nor with scattering on the impurity) between left and right electrodes. Spin current can then flow through the molecule in a biased system due to single-electrode tunneling processes, and thus this spin current is not associated with any charge transport. Interestingly, it has been shown that in the case of an asymmetric junction, where the asymmetry is either due to different DOSs at the Fermi level in both electrodes or due to different exchange coupling of the impurity with electrodes, the mechanism of spin transport under discussion can be  more efficient than spin transport associated with charge transfer between electrodes in a conventional magnetic tunnel junction, i.e. without a spin impurity.

\acknowledgments

M.M. thanks the Kavli Institute for Theoretical Physics at University of California in Santa Barbara for hospitality.
This work was supported by the National Science Center in Poland as the Project No. DEC-2012/04/A/ST3/00372.
\mbox{M.\,M.} acknowledges support from  the Alexander von Humboldt Foundation. This research was  supported in part by the National Science Foundation under Grant No. PHY11-25915.

\appendix
\onecolumngrid
\section{\label{App:kin_coeff}Expressions for currents}
Equations~(\ref{Eq:I_C_def})-(\ref{Eq:I_Q_def}) can be written in a more explicit form. After changing summations with respect
to wave vectors to integration over energy one obtains
\begingroup
\allowdisplaybreaks
    \begin{align}
    \label{Eq:App_A_I_C}
    &I_\text{C}
    =e\frac{2\pi}{\hbar}
    \sum_{\alpha\beta}
    \sum_{\chi\chi^\prime}
    \big|\bra{R\beta, \chi^\prime}\mathcal{H}_\textrm{int}\ket{L\alpha,\chi}\big|^2
    \int\textrm{d}\omega\
    \rho_\alpha^L(\omega)
    \rho_\beta^R(\omega+\Delta_{\chi\chi^\prime})
                        \nonumber\\[-5pt]
    &\hspace*{150pt}
    \times
    \Big\{
    \mathcal{P}_\chi
    f_{L\alpha}\big(\omega\big)
    \Big[
    1-f_{R\beta}\big(\omega+\Delta_{\chi\chi^\prime}\big)
    \Big]
    -
    \mathcal{P}_{\chi^\prime}
    f_{R\beta}\big(\omega+\Delta_{\chi\chi^\prime}\big)
    \Big[
    1-f_{L\alpha}\big(\omega\big)
    \Big]
    \Big\}
    ,
                    \\[5pt]
    & I_\text{S}=
    \frac{\hbar}{2}
    \frac{2\pi}{\hbar}
    \sum_{\chi\chi^\prime}
    \Bigg\{
    \sum_{\alpha}
    \eta_\alpha
    \big|\bra{R\alpha, \chi^\prime}\mathcal{H}_\textrm{int}\ket{L\alpha,\chi}\big|^2
    \Bigg[
    \mathcal{P}_\chi
     \int\textrm{d}\omega
    \rho_\alpha^L(\omega)
    \rho_\alpha^R(\omega+\Delta_{\chi\chi^\prime})
    f_{L\alpha}\big(\omega\big)
    \Big[1-f_{R\alpha}\big(\omega+\Delta_{\chi\chi^\prime}\big)\Big]
                    \nonumber\\[-8pt]
    &\hspace*{187pt}-\mathcal{P}_{\chi^\prime}
    \int\textrm{d}\omega
    \rho_\alpha^L(\omega)
    \rho_\alpha^R(\omega+\Delta_{\chi\chi^\prime})
    f_{R\alpha}\big(\omega+\Delta_{\chi\chi^\prime}\big)
    \Big[1-f_{L\alpha}\big(\omega\big)\Big]
    \Bigg]
                    \nonumber\\[-5pt]
    &\hspace*{60pt}+\sum_q
    \eta_q
    \big|\bra{q\downarrow, \chi^\prime}\mathcal{H}_\textrm{int}\ket{q\uparrow,\chi}\big|^2
    \Bigg[
    \mathcal{P}_\chi
     \int\textrm{d}\omega
    \rho_\uparrow^q(\omega)
    \rho_\downarrow^q(\omega+\Delta_{\chi\chi^\prime})
    f_{q\uparrow}\big(\omega\big)
    \Big[1-f_{q\downarrow}\big(\omega+\Delta_{\chi\chi^\prime}\big)\Big]
                    \nonumber\\[-8pt]
    &\hspace*{184pt}-\mathcal{P}_{\chi^\prime}
    \int\textrm{d}\omega
    \rho_\uparrow^q(\omega)
    \rho_\downarrow^q(\omega+\Delta_{\chi\chi^\prime})
    f_{q\downarrow}\big(\omega+\Delta_{\chi\chi^\prime}\big)
    \Big[1-f_{q\uparrow}\big(\omega\big)\Big]
    \Bigg]
    \Bigg\}
    ,
                    \\[5pt]
    \label{Eq:App_A_I_Q}
    &I_\text{Q}=
    \frac{2\pi}{\hbar}
    \sum_{\alpha\beta}
    \sum_{\chi\chi^\prime}
    \big|\bra{R\beta, \chi^\prime}\mathcal{H}_\textrm{int}\ket{L\alpha,\chi}\big|^2
    \int\text{d}\omega
    \rho_\alpha^L(\omega)
    \rho_\beta^R(\omega+\Delta_{\chi\chi^\prime})
                    \nonumber\\[-8pt]
    &\hspace*{170pt}\times
    \Bigg\{
    \mathcal{P}_\chi
    \Big(
    \omega+\tfrac{1}{2}\Delta_{\chi\chi^\prime}-\mu_0-\tfrac{1}{2}\big(\eta_\alpha-\eta_\beta\big)e\delta V_\text{S}
    \Big)
    f_{L\alpha}\big(\omega\big)
    \Big[1-f_{R\beta}\big(\omega+\Delta_{\chi\chi^\prime}\big)\Big]
                    \nonumber\\[-5pt]
    &\hspace*{178pt}-
    \mathcal{P}_{\chi^\prime}
    \Big(
    \omega+\tfrac{1}{2}\Delta_{\chi\chi^\prime}-\mu_0-\tfrac{1}{2}\big(\eta_\alpha-\eta_\beta\big)e\delta V_\text{S}
    \Big)
    f_{R\beta}\big(\omega+\Delta_{\chi\chi^\prime}\big)
    \Big[1-f_{L\alpha}\big(\omega\big)\Big]
    \Bigg\}
                     \nonumber\\[-5pt]
    &\hspace*{12pt}-\frac{2\pi}{\hbar}
    \sum_q
    \sum_{ \alpha\beta}
    \sum_{\chi\chi^\prime}
    \eta_q
    \mathcal{P}_\chi
    \big|\bra{q\beta, \chi^\prime}\mathcal{H}_\textrm{int}\ket{q\alpha,\chi}\big|^2
    \int\text{d}\omega
    \rho_\alpha^q(\omega)
    \rho_\beta^q(\omega+\Delta_{\chi\chi^\prime})
                    \nonumber\\[-5pt]
    &\hspace*{210pt}\times
    \Big(
    \tfrac{1}{2}\Delta_{\chi\chi^\prime}+\tfrac{1}{4}\eta_q\big(\eta_\alpha-\eta_\beta\big)e\delta V_\text{S}
    \Big)
    f_{q\alpha}\big(\omega\big)
    \Big[1-f_{q\beta}\big(\omega+\Delta_{\chi\chi^\prime}\big)\Big]
    .
    \end{align}
\endgroup
%
\section{\label{App:kin_coeff}Derivation of the kinetic coefficients}

In order to derive the linear-response expressions  for the charge ($I_\text{C}\equiv I_0$), spin ($I_\text{S}\equiv I_1$)
and heat ($I_\text{Q}\equiv I_2$) currents, one has to linearize them  with respect to the voltage bias   $\delta V$,
spin bias $\delta V_\text{S}$ as well as temperature difference~$\delta T$,
    \begin{equation}\label{Eq:I_n_lin_resp_expand}
    I_n
    \approx
    \frac{\partial I_n}{\partial \delta V}\Big|_\text{eq}
    \delta V
    +
    \frac{\partial I_n}{\partial \delta V_\text{S}}\Big|_\text{eq}
    \delta V
    +
    \frac{\partial I_n}{\partial \delta T}\Big|_\text{eq}
    \delta T,
    \end{equation}
for $n=0,1,2$, where the subscript 'eq' means equilibrium situation, i.e. $\delta V=\delta V_S=\delta T =0$.
According to Eq.~(\ref{Eq:I_n_matrix_kin_coeff}) the formulae for the charge, spin and heat currents can be expressed in
terms of the kinetic coefficients $\mathcal{L}_{nk}$ as
    \begin{equation}
    I_n=
    e^{\delta_{n0}}
    \Big(
    \frac{\hbar}{2}
    \Big)^{\!\delta_{n1}}
    \Big[
    e\mathcal{L}_{n0}x_0
    +
    e\mathcal{L}_{n1}x_1
    +
    \frac{1}{T}\mathcal{L}_{n2}x_2
    \Big].
    \end{equation}
In the  equation above, the shorthand notation $x_0\equiv \delta V$, $x_1\equiv \delta V_\text{S}$ and $x_2\equiv \delta T$ has been introduced.

From Eq.~(\ref{Eq:I_n_lin_resp_expand}), it becomes clear that in order to derive the kinetic coefficients one has to
calculate the relevant derivatives of Eqs.~(\ref{Eq:I_C_def})-(\ref{Eq:I_Q_def}) [or essentially its integral versions
given by Eqs.~(\ref{Eq:App_A_I_C})-(\ref{Eq:App_A_I_Q})], which in general is a nontrivial task since it requires the
knowledge of the explicit form of the probabilities $\mathcal{P}_\chi$.
Since we are interested in the linear-response regime and \emph{elastic} contributions to transport processes,
we take into account only degenerate
ground states if their energy separation from excited states is significantly larger than thermal energy.
For an \emph{isotropic} spin impurity of arbitrary spin number $S$, the analytical solution can be then
found without any further approximations regarding the impurity spectrum (see Sec.~\ref{App:P_isotropic}). On the other
hand, for an \emph{anisotropic} half-integer spin impurity ($S>1/2$) exhibiting magnetic anisotropy, the problem can be
significantly simplified for sufficiently low temperatures, $T\ll(2S-1)D$ by truncating the impurity spectrum to the ground
state Kramers' dublet, so that only transitions within the doublet  $\ket{\chi_S}$ and $\ket{\chi_{-S}}$ can
occur (see Sec.~2 of this Appendix).  One can find the kinetic coefficients to have the general form
    \begin{align}\label{Eq:kin_coeff_general}
    \hspace*{-5pt}
     \mathcal{L}_{nk}=
     \frac{2\pi}{\hbar}K^2
     \sum_{\chi\chi^\prime}
     \Bigg[
     \frac{T^{\delta_{k2}}}{e^{\delta_{k0}+\delta_{k1}}}
     \Bigg(
     &
    \sum_{\sigma\sigma^\prime}
    (\eta_\sigma\delta_{\sigma\sigma^\prime})^{\delta_{n1}}
    \mathcal{W}_{\sigma\sigma^\prime}^{\chi\chi^\prime}
    \int\textrm{d}\omega\
    \rho_\sigma^L(\omega)
    \rho_{\sigma^\prime}^R(\omega)
     \Bigg\{
     \Bigg(
    \frac{\partial \mathcal{P}_\chi}{\partial x_k}\Big|_\text{eq}
     -
      \frac{\partial \mathcal{P}_{\chi^\prime}}{\partial x_k}\Big|_\text{eq}
     \Bigg)
     f\big(\omega\big)
    \Big[
    1-f\big(\omega\big)
    \Big]
                    \nonumber\\
     &\hspace*{10pt}+
     \mathcal{P}_\chi\Big|_\text{eq}
     \,
     \frac{\partial}{\partial x_k}
    \Bigg(
    f_{L\sigma}\big(\omega\big)
    \Big[
    1-f_{R\sigma^\prime}\big(\omega\big)
    \Big]
    \Bigg)\Big|_\text{eq}
    -
     \mathcal{P}_{\chi^\prime}\Big|_\text{eq}
     \,
     \frac{\partial}{\partial x_k}
    \Bigg(
    \Big[
    1-f_{L\sigma}\big(\omega\big)
    \Big]
    f_{R\sigma^\prime}\big(\omega\big)
    \Bigg)\Big|_\text{eq}
    \Bigg\}
                            \nonumber\\
    +
    \delta_{n1}
    &
    \sum_q
    \eta_q
    \big(\alpha_\text{ex}^{qq}\big)^{\!2}
    \big|\mathbb{S}_{\chi^\prime\chi}^+\big|^2
    \int\textrm{d}\omega\
    \rho_\uparrow^q(\omega)
    \rho_\downarrow^q(\omega)
    \Bigg\{
     \Bigg(
    \frac{\partial \mathcal{P}_\chi}{\partial x_k}\Big|_\text{eq}
     -
      \frac{\partial \mathcal{P}_{\chi^\prime}}{\partial x_k}\Big|_\text{eq}
     \Bigg)
     f\big(\omega\big)
    \Big[
    1-f\big(\omega\big)
    \Big]
                            \nonumber\\
     &\hspace*{10pt}+
     \mathcal{P}_\chi\Big|_\text{eq}
     \,
     \frac{\partial}{\partial x_k}
    \Bigg(
    f_{q\uparrow}\big(\omega\big)
    \Big[
    1-f_{q\downarrow}\big(\omega\big)
    \Big]
    \Bigg)\Big|_\text{eq}
    -
     \mathcal{P}_{\chi^\prime}\Big|_\text{eq}
     \,
     \frac{\partial}{\partial x_k}
    \Bigg(
    \Big[
    1-f_{q\uparrow}\big(\omega\big)
    \Big]
    f_{q\downarrow}\big(\omega\big)
    \Bigg)\Big|_\text{eq}
    \Bigg\}
    \Bigg)
                    \nonumber\\
    -
    \delta_{n2}\delta_{k1}
    \frac{1}{2}
    &
    \sum_q
    \sum_{\sigma}
    \big(\alpha_\text{ex}^{qq}\big)^{\!2}
    \Big(
    \delta_{\sigma\uparrow}
    \big|\mathbb{S}_{\chi^\prime\chi}^+\big|^2
    -
    \delta_{\sigma\downarrow}
    \big|\mathbb{S}_{\chi^\prime\chi}^-\big|^2
    \Big)
    \mathcal{P}_\chi\Big|_\text{eq}
    \int\textrm{d}\omega\
    \rho_\sigma^q(\omega)
    \rho_{\overline{\sigma}}^q(\omega)
     f\big(\omega\big)
    \Big[
    1-f\big(\omega\big)
    \Big]
    \Bigg]
    ,
    \end{align}
where the summation for isotropic case is over all spin state, while for anisotropic case over two degenerate states of lowest energy, and
    \begin{align}
    \mathcal{W}_{\sigma\sigma^\prime}^{\chi\chi^\prime}
    =
    \delta_{\sigma^\prime\sigma}
    \delta_{\chi^\prime\chi}\,
    \Big[
    \alpha_\text{d}^2
    +
    2\eta_\sigma
    \alpha_\text{d}
    \alpha_\text{ex}^{LR}\,
    \mathbb{S}_{\chi\chi}^z
    \Big]
    +
    \big(\alpha_\text{ex}^{LR}\big)^{2}
    \Big\{
    \delta_{\sigma^\prime\sigma}
    \big|\mathbb{S}_{\chi^\prime\chi}^z\big|^2
    \!
    +
    \!
    \delta_{\sigma^\prime\overline{\sigma}}\,
    \Big[
    \delta_{\sigma\downarrow}
    \big|\mathbb{S}_{\chi^\prime\chi}^-\big|^2
    \!
    +
    \!
    \delta_{\sigma\uparrow}
    \big|\mathbb{S}_{\chi^\prime\chi}^+\big|^2
    \Big]
    \Big\}
    .
    \end{align}
To show that Eq.~(\ref{Eq:kin_coeff_general}) indeed represents the kinetic coefficient, one should additionally prove
that it satisfies the Onsager relation. For  this purpose, let's begin with calculating the derivatives involving the
Fermi-Dirac distribution functions of electrodes,
    \begin{align}
    \frac{\partial}{\partial x_k}
    \Bigg(
    f_{q\sigma}(\omega)\Big[1-f_{q^\prime\sigma^\prime}(\omega)\Big]
    \Bigg)\!\Big|_\text{eq}
    =
    \frac{\partial f_{q\sigma}(\omega)}{\partial x_k}\Big|_\text{eq}
    -
    \Bigg(
    \frac{\partial f_{q\sigma}(\omega)}{\partial x_k}\Big|_\text{eq}
    +
    \frac{\partial f_{q^\prime\sigma^\prime}(\omega)}{\partial x_k}\Big|_\text{eq}
    \Bigg)
     f(\omega)
     ,
    \end{align}
where we used that $f_{q\sigma}(\omega)\big|_\text{eq}\equiv f(\omega)=\big\{1+\exp\big[(\omega-\mu_0)T^{-1}\big]\!\big\}^{-1}$, and
    \begin{align}\label{Eq:FD_fun_deriv}
    \hspace*{-1pt}
    \frac{\partial f_{q\sigma}(\omega)}{\partial x_k}\Big|_\text{eq}
    \!\!
    =
    \eta_q
    \eta_\sigma^{\delta_{k1}}
    &
    \frac{e^{\delta_{k0}+\delta_{k1}}}{2T^{\delta_{k2}}}
    \big(\omega-\mu_0\big)^{\delta_{k2}}
    \Big[
    \!
    -
    f^\prime\!(\omega)
    \Big],
    \end{align}
with  $f^\prime\!(\omega)\equiv\partial f(\omega)/\partial\omega$.
Next, we define an auxiliary function
    \begin{align}\label{Eq:aux_phi_function}
    \phi_{\sigma\sigma^\prime}^{(n,k)qq^\prime}
    =
    \int\!\text{d}\omega
    \rho_\sigma^q(\omega)
    \rho_{\sigma^\prime}^{q^\prime}(\omega)
    \big(\omega-\mu_0\big)^n
    \Big[
    \!
    -
    f^\prime\!(\omega)
    \Big]
    \big[
    f(\omega)
    \big]^k,
    \end{align}
obeying the symmetry relation $\phi_{\sigma\sigma^\prime}^{(n,k)qq^\prime}=\phi_{\sigma^\prime\sigma}^{(n,k)q^\prime q}$,
and we employ the identity $f(\omega)\big[1-f(\omega)\big]=T\big[-f^\prime\!(\omega)\big]$, which in consequence allows for writing
    \begin{align}\label{Eq:kin_coeff_app}
     \mathcal{L}_{nk}=\frac{\pi}{\hbar}
     K^2
     \sum_{\chi\chi^\prime}
     \Bigg[
     &
    \sum_{\sigma\sigma^\prime}
    (\eta_\sigma\delta_{\sigma\sigma^\prime})^{\delta_{n1}}
    \mathcal{W}_{\sigma\sigma^\prime}^{\chi\chi^\prime}
     \Bigg\{
     \Bigg(
     \eta_\sigma^{\delta_{k1}}
     \mathcal{P}_\chi\Big|_\text{eq}
     +
     \eta_{\sigma^\prime}^{\delta_{k1}}
      \mathcal{P}_{\chi^\prime}\Big|_\text{eq}
     \Bigg)
    \phi_{\sigma\sigma^\prime}^{(\delta_{n2}+\delta_{k2},0)LR}
                                    \nonumber\\
    &\hspace*{220pt}
    +
     \frac{2T^{1+\delta_{k2}}}{e^{\delta_{k0}+\delta_{k1}}}
     \Bigg(
    \frac{\partial \mathcal{P}_\chi}{\partial x_k}\Big|_\text{eq}
     -
      \frac{\partial \mathcal{P}_{\chi^\prime}}{\partial x_k}\Big|_\text{eq}
     \Bigg)
     \phi_{\sigma\sigma^\prime}^{(\delta_{n2},0)LR}
    \Bigg\}
                            \nonumber\\
    +\ &
    \delta_{n1}
    \sum_q
    \eta_q
     \big(\alpha_\text{ex}^{qq}\big)^{\!2}
    \big|\mathbb{S}_{\chi^\prime\chi}^+\big|^2
    \Bigg\{
    \eta_q
     \Bigg(
     \mathcal{P}_\chi\Big|_\text{eq}
     -
    (-1)^{\delta_{k1}}
     \mathcal{P}_{\chi^\prime}\Big|_\text{eq}
     \Bigg)
    \phi_{\uparrow\downarrow}^{(\delta_{k2},0)qq}
                          \nonumber\\
    &\hspace*{220pt}
    +
     \frac{2T^{1+\delta_{k2}}}{e^{\delta_{k0}+\delta_{k1}}}
     \Bigg(
    \frac{\partial \mathcal{P}_\chi}{\partial x_k}\Big|_\text{eq}
    -
     \frac{\partial \mathcal{P}_{\chi^\prime}}{\partial x_k}\Big|_\text{eq}
    \Bigg)
     \phi_{\uparrow\downarrow}^{(0,0)qq}
    \Bigg\}
                    \nonumber\\
     -\ &
    \delta_{n2}\delta_{k1}
    T
    \sum_q
    \sum_{\sigma}
    \big(\alpha_\text{ex}^{qq}\big)^{\!2}
    \Big(
    \delta_{\sigma\uparrow}
    \big|\mathbb{S}_{\chi^\prime\chi}^+\big|^2
    -
    \delta_{\sigma\downarrow}
    \big|\mathbb{S}_{\chi^\prime\chi}^-\big|^2
    \Big)
     \mathcal{P}_\chi\Big|_\text{eq}
     \phi_{\sigma\overline{\sigma}}^{(0,0)qq}
     \Bigg]
     .
    \end{align}
We point out that the equations above have been simplified  by noting that
$\mathcal{P}_\chi\big|_\text{eq}-\mathcal{P}_{\chi^\prime}\big|_\text{eq}=0$ (i.e. at equilibrium)
for any pair of degenerate states $\ket{\chi}$ and $\ket{\chi^\prime}$ [for details see the discussion below].

In the next step  we
have to find the linear coefficients of the Taylor expansion for the probabilities  with respect to  voltage
and spin bias, as well as to temperature difference. We discuss the derivation procedure separately for the case
of an \emph{isotropic} spin impurity (Sec.~\ref{App:P_isotropic})  and an \emph{anisotropic} spin impurity
with magnetic anisotropy (Sec.~\ref{App:P_anisotropic}), where only the Kramers' ground doublet is taken into consideration.

\subsection{\label{App:P_isotropic}Isotropic spin impurity $S$}

As described in Sec.~\ref{Sec:Transport_charcter}, the probabilities of finding the impurity in a specific spin state
can be found by means of the set of stationary master equations, see Eq.~(\ref{Eq:master_eqs}). Because  the  Hamiltonian
of the impurity is rotationally invariant in the present case, one can conveniently use the eigenvalues $m$ of the $z$th
component of the spin operator $S_z$ to label the spin states, i.e. $\ket{\chi}\equiv\ket{m}$, so that
    \begin{equation}\label{Eq:master_eq_NOanisotrop}
    \underset{m}{\forall}\
    \mathcal{P}_{m-1}\gamma_{m-1,m}
    +
    \mathcal{P}_{m+1}\gamma_{m+1,m}
    -
    \mathcal{P}_{m}\big(\gamma_{m,m-1}+\gamma_{m,m+1}\big)
    =
    0,
    \end{equation}
with transition rates given by
    \begin{equation}\label{Eq:gamma_def_NOanisotrop}
    \left\{
    \begin{aligned}
    &\gamma_{m,m-1}=
    \frac{2\pi}{\hbar}
    K^2
    \mathcal{C}_m^-
    \sum_{qq^\prime}
    \big(\alpha_\text{ex}^{qq^\prime}\big)^{\!2}
    \Phi_{\downarrow\uparrow}^{(0)qq^\prime}\!(0),
    \\
    &\gamma_{m-1,m}=
    \frac{2\pi}{\hbar}
    K^2
    \mathcal{C}_m^-
    \sum_{qq^\prime}
    \big(\alpha_\text{ex}^{qq^\prime}\big)^{\!2}
    \Phi_{\uparrow\downarrow}^{(0)qq^\prime}\!(0),
    \end{aligned}
    \right.
    \quad
    \text{with}
    \quad
    \mathcal{C}_m^\pm
    \equiv
    S(S+1)-m(m\pm1)
    .
    \end{equation}
Using Eq.~(\ref{Eq:master_eq_NOanisotrop}) together with the probability normalization condition $\sum_m\mathcal{P}_m=1$, one obtains
    \begin{equation}
    \mathcal{P}_S
    =
    \Bigg[
    1
    +
    \sum_{m=0}^{2S-1}\prod_{k=0}^m
    \frac{\gamma_{S-k,S-1-k}}{\gamma_{S-1-k,S-k}}
    \Bigg]^{-1}
    \quad
    \text{and}
    \quad
    \underset{m\neq S}{\forall}\
    \mathcal{P}_m
    =
    \mathcal{P}_S
    \prod_{k=0}^{S-1-m}
    \frac{\gamma_{m+1+k,m+k}}{\gamma_{m+k,m+1+k}},
    \end{equation}
and after taking into account  Eq.~(\ref{Eq:gamma_def_NOanisotrop}),
    \begin{equation}
    \mathcal{P}_m
    =
    \frac{
    \mathcal{Y}^{S-m}
    }{
    1
    +
    \sum_{m=0}^{2S-1}
    \mathcal{Y}^{m+1}
    }
    \quad
    \text{with}
    \quad
    \mathcal{Y}
    \equiv
    \dfrac{
    \sum_{qq^\prime}
    \big(\alpha_\text{ex}^{qq^\prime}\big)^{\!2}
    \Phi_{\downarrow\uparrow}^{(0)qq^\prime}\!(0)
    }{
    \sum_{qq^\prime}
    \big(\alpha_\text{ex}^{qq^\prime}\big)^{\!2}
    \Phi_{\uparrow\downarrow}^{(0)qq^\prime}\!(0)
    }
    .
    \end{equation}

By noting that at equilibrium  the function $\Phi_{\sigma\sigma^\prime}^{(n)qq^\prime}\!\!(0)$, see Eq.~(\ref{Eq:Phi_function_def}),
exhibits the following symmetry property
    \begin{equation}\label{Eq:Phi_function_sym_property}
    \Phi_{\sigma\sigma^\prime}^{(n)qq^\prime}\!\!(0)\Big|_\text{eq}
    =
    \Phi_{\sigma^\prime\!\sigma}^{(n)q^\prime\!q}(0)\Big|_\text{eq}
    ,
    \end{equation}
 one obtains $\mathcal{Y}\big|_\text{eq}=1$, which allows for writing the equilibrium probabilities and the corresponding
 derivatives in the following form
    \begin{equation}
    \mathcal{P}_m\Big|_\text{eq}
    =
    \frac{1}{2S+1}
    \quad
    \text{and}
    \quad
    \frac{\partial \mathcal{P}_m}{\partial x_k}\Big|_\text{eq}
    =
     -m
    \frac{\partial \mathcal{Y}}{\partial x_k}\Big|_\text{eq}
    =
    -m\,
    \dfrac{
    \sum\limits_{qq^\prime}
    \big(\alpha_\text{ex}^{qq^\prime}\big)^{\!2}
    \Bigg[
    \dfrac{\partial \Phi_{\downarrow\uparrow}^{(0)qq^\prime}\!(0)}{\partial x_k}
    \Big|_\text{eq}
    -
    \dfrac{\partial \Phi_{\uparrow\downarrow}^{(0)qq^\prime}\!(0)}{\partial x_k}
    \Big|_\text{eq}
    \Bigg]
    }{
    \sum\limits_{qq^\prime}
    \big(\alpha_\text{ex}^{qq^\prime}\big)^{\!2}
    \Phi_{\uparrow\downarrow}^{(0)qq^\prime}(0)\big|_\text{eq}
    }
    .
    \end{equation}
Finally, employing that
    \begin{align}\label{Eq:Phi_deriv}
    \frac{\partial \Phi_{\sigma\overline{\sigma}}^{(0)qq^\prime}(0)}{\partial x_k}\Big|_\text{eq}
    =
    \frac{e^{\delta_{k0}+\delta_{k1}}}{2T^{\delta_{k2}}}
    \Big[
    \eta_q
    \eta_\sigma^{\delta_{k1}}
    \phi_{\sigma\overline{\sigma}}^{(\delta_{k2},0)qq^\prime}
    -
    (\eta_\sigma^{\delta_{k1}}\eta_q+\eta_{\overline{\sigma}}^{\delta_{k1}}\eta_{q^\prime})
    \phi_{\sigma\overline{\sigma}}^{(\delta_{k2},1)qq^\prime}
    \Big]
    ,
    \end{align}
we get the explicit expression for the difference of probabilities' derivatives
    \begin{align}
    \frac{\partial \mathcal{P}_m}{\partial x_k}\Big|_\text{eq}
    -
    \frac{\partial \mathcal{P}_{m^\prime}}{\partial x_k}\Big|_\text{eq}
    =\ &
    \frac{m-m^\prime}{2S+1}
    \Omega
    \frac{e^{\delta_{k0}+\delta_{k1}}}{T^{\delta_{k2}}}
    \Bigg\{
    \big(\delta_{k0}+\delta_{k2}\big)
    \big(\alpha_\text{ex}^{LR}\big)^{\!2}
    \sum_\sigma
    \eta_\sigma
    \phi_{\sigma\overline{\sigma}}^{(\delta_{k2},0)LR}
    +
    \delta_{k1}
    \sum_q
    \eta_q
    \big(\alpha_\text{ex}^{qq}\big)^{\!2}
    \phi_{\uparrow\downarrow}^{(\delta_{k2},0)qq}
    \Bigg\}
    ,
    \end{align}
with
    \begin{equation}
    \Omega
    =
    \Bigg[
    \sum_{qq^\prime}
    \big(\alpha_\text{ex}^{qq^\prime}\big)^2
    \Phi_{\uparrow\downarrow}^{(0)qq^\prime}\!(0)\Big|_\text{eq}
    \Bigg]^{-1}
    .
    \end{equation}
Next, by noting that for $m\neq m^\prime$ holds
    \begin{align}
    \mathcal{W}_{\sigma\sigma^\prime}^{mm^\prime}
    =
    \big(\alpha_\text{ex}^{LR}\big)^{\!2}
    \delta_{\sigma^\prime\overline{\sigma}}
    \Big\{
    \delta_{\sigma\downarrow}
    \delta_{m,m^\prime+1}
    \mathcal{C}_{m^\prime}^+
    +
    \!
    \delta_{\sigma\uparrow}
    \delta_{m^\prime\!,m+1}
    \mathcal{C}_m^+
    \Big\}
    ,
    \end{align}
and $\sum_m\mathcal{C}_m^+=(2/3)S(S+1)(2S+1)$,
we arrive at the final formula for the kinetic coefficient:
    \begin{align}
    \hspace*{-5pt}
     \mathcal{L}_{nk}=\frac{\pi}{\hbar}
     K^2
     \Bigg[
     &
     \frac{2}{2S+1}
     \sum_{mm^\prime}
    \sum_{\sigma\sigma^\prime}
    (\eta_\sigma\delta_{\sigma\sigma^\prime})^{\delta_{n1}+\delta_{k1}}
    \mathcal{W}_{\sigma\sigma^\prime}^{mm^\prime}
    \phi_{\sigma\sigma^\prime}^{(\delta_{n2}+\delta_{k2},0)LR}
                            \nonumber\\
    -\ &
    \big(\delta_{n0}+\delta_{n2}\big)
    \big(\delta_{k0}+\delta_{k2}\big)
    \frac{4}{3}
    S(S+1)
    \Omega
    T
    \big(\alpha_\text{ex}^{LR}\big)^{\!4}
    \sum_{\sigma\sigma^\prime}
    \eta_\sigma
    \eta_{\sigma^\prime}
    \phi_{\sigma\overline{\sigma}}^{(\delta_{n2},0)LR}
    \phi_{\sigma^\prime\overline{\sigma^\prime}}^{(\delta_{k2},0)LR}
                              \nonumber\\
    -\ &
    \Big[
    \big(\delta_{n0}+\delta_{n2}\big)
    \delta_{k1}
    +
    \delta_{n1}
    \big(\delta_{k0}+\delta_{k2}\big)
    \Big]
    \frac{4}{3}
    S(S+1)
    \Omega
    T
    \big(\alpha_\text{ex}^{LR}\big)^{\!2}
    \sum_{q\sigma}
    \eta_q
    \eta_\sigma
     \big(\alpha_\text{ex}^{qq}\big)^{\!2}
    \phi_{\sigma\overline{\sigma}}^{(\delta_{n2}+\delta_{k2},0)LR}
    \phi_{\uparrow\downarrow}^{(0,0)qq}
                            \nonumber\\
    -\ &
    \delta_{n1}
    \delta_{k1}
    \Bigg\{
    \frac{4}{3}
    S(S+1)
    \Omega
    T
    \sum_{qq^\prime}
    \eta_q
    \eta_{q^\prime}
     \big(\alpha_\text{ex}^{qq}\big)^{\!2}
     \big(\alpha_\text{ex}^{q^\prime\! q^\prime}\big)^{\!2}
     \phi_{\uparrow\downarrow}^{(0,0)qq}
     \phi_{\uparrow\downarrow}^{(\delta_{k2},0)q^\prime\! q^\prime}
     -
     \frac{4}{3}
    S(S+1)
    \sum_q
     \big(\alpha_\text{ex}^{qq}\big)^{\!2}
    \phi_{\uparrow\downarrow}^{(0,0)qq}
     \Bigg\}
     \Bigg]
     .
     \hspace*{-3pt}
    \end{align}
It can be easily seen that the expression above is symmetric with respect to exchanging the indices $n$ and $k$, and hence it satisfies the Onsager relation.

\subsection{\label{App:P_anisotropic}Anisotropic spin impurity $S>1/2$}

A similar derivation as above can be performed for the case of an anisotropic spin impurity  with both uniaxial and
transverse component of magnetic anisotropy. Provided that only the  Kramers' doublet states $\ket{\chi_S}$ and $\ket{\chi_{-S}}$ participate
in the transport, the general expression for the probabilities of finding the impurity spin in either of these two states can be easily
found from Eq.~(\ref{Eq:master_eqs}) and the normalization condition for probability,
    \begin{equation}\label{eq:probab_GS_def}
    \mathcal{P}_{\chi_S}=\frac{\gamma_{\chi_{-S}\chi_{S}}}{\gamma}
    \quad
    \textrm{and}
    \quad
    \mathcal{P}_{\chi_{-S}}=\frac{\gamma_{\chi_{S}\chi_{-S}}}{\gamma}
    \end{equation}
with $\gamma=\gamma_{\chi_{-S}\chi_{S}}+\gamma_{\chi_{S}\chi_{-S}}$. From Eq.~(\ref{Eq:trans_rates_general}) one gets for $m=\pm S$
    \begin{align}\label{Eq:gamma_chiMchi-M}
    \gamma_{\chi_m\chi_{-m}}
    =
    \frac{2\pi}{\hbar}
    K^2
    \sum_{qq^\prime\sigma}
    \big(\alpha_\text{ex}^{qq^\prime}\big)^{\!2}\,
    \Phi_{\sigma\overline{\sigma}}^{(0)qq^\prime}\!\!(0)
    \Big[
    \delta_{\sigma\downarrow}
    \big|\mathbb{S}_{\chi_{-m}\chi_m}^-\big|^2
    +
    \delta_{\sigma\uparrow}
    \big|\mathbb{S}_{\chi_{-m}\chi_m}^+\big|^2
    \Big]
    ,
    \end{align}
and
    \begin{equation}
    \hspace*{-2pt}
    \gamma
    =
    \frac{2\pi}{\hbar}
    K^2
    \!
    \sum_{\xi=\pm}
    \!
    \big|\mathbb{S}_{\chi_{-S}\chi_{S}}^\xi\big|^2
    \!
    \sum_{qq^\prime\sigma}
    \!
    \big(\alpha_\text{ex}^{qq^\prime}\big)^{\!2}\,
    \Phi_{\sigma\overline{\sigma}}^{(0)qq^\prime}\!\!(0)
    .
    \end{equation}
Using the symmetry property of $\Phi_{\sigma\sigma^\prime}^{(n)qq^\prime}\!\!(0)\big|_\text{eq}$, see Eq.~(\ref{Eq:Phi_function_sym_property}),
one can show that
    \begin{equation}\label{Eq:trans_rates_GS_eq}
    \gamma_{\chi_S\chi_{-S}}\Big|_\text{eq}
    =
    \gamma_{\chi_{-S}\chi_{S}}\Big|_\text{eq}
    =
    \frac{1}{2}
    \gamma\Big|_\text{eq}
    \quad
    \text{and, consequently,}
    \quad
    \mathcal{P}_{\chi_S}\Big|_\text{eq}
    =
    \mathcal{P}_{\chi_{-S}}\Big|_\text{eq}
    =
    \frac{1}{2}
    .
    \end{equation}
Furthermore, employing Eqs.~(\ref{eq:probab_GS_def}) and~(\ref{Eq:trans_rates_GS_eq}), one finds for $m=\pm S$
    \begin{align}\label{Eq:P_deriv}
    \frac{\partial \mathcal{P}_{\chi_m}}{\partial x_k}\Big|_\text{eq}
    =
    \frac{1}{\gamma\big|_\text{eq}}
    \Bigg[
    \frac{\partial \gamma_{\chi_{-m}\chi_m}}{\partial x_k}\Big|_\text{eq}
    -
    \frac{1}{2}
    \frac{\partial \gamma}{\partial x_k}\Big|_\text{eq}
    \Bigg]
    ,
    \end{align}
from where it immediately follows that
    \begin{align}
    \frac{\partial \mathcal{P}_{\chi_m}}{\partial x_k}\Big|_\text{eq}
    -
    \frac{\partial \mathcal{P}_{\chi_{-m}}}{\partial x_k}\Big|_\text{eq}
    =
    \frac{1}{\gamma\big|_\text{eq}}
    \Bigg[
    \frac{\partial \gamma_{\chi_{-m}\chi_m}}{\partial x_k}\Big|_\text{eq}
    -
    \frac{\partial \gamma_{\chi_{m}\chi_{-m}}}{\partial x_k}\Big|_\text{eq}
    \Bigg],
    \end{align}
Using then Eq.~(\ref{Eq:Phi_deriv}), one obtains
    \begin{align}\label{Eq:derivPdiff}
    \frac{\partial \mathcal{P}_{\chi_m}}{\partial x_k}\Big|_\text{eq}
    -
    \frac{\partial \mathcal{P}_{\chi_{-m}}}{\partial x_k}\Big|_\text{eq}
    =
    -
    \frac{1}{2}\,
    \text{sgn}_z(\chi_m)
    \frac{\Lambda_-}{\Lambda_+}
    \Omega
    \frac{e^{\delta_{k0}+\delta_{k1}}}{T^{\delta_{k2}}}
    \Bigg\{
    &
    (
    \delta_{k0}
    +
    \delta_{k2}
    )
    \big(\alpha_\text{ex}^{LR}\big)^{\!2}
    \sum_\sigma
    \eta_\sigma
    \phi_{\sigma\overline{\sigma}}^{(\delta_{k2},0)LR}
                        \nonumber\\
    +\ &
     \delta_{k1}
    \sum_q
    \eta_q
    \big(\alpha_\text{ex}^{qq}\big)^{\!2}
    \phi_{\uparrow\downarrow}^{(\delta_{k2},0)qq}
    \Bigg\}
    ,
    \end{align}
where $\text{sgn}_z(\chi_m)\equiv\text{sgn}\big(\mathbb{S}_{\chi_m\chi_m}^z\big)$ and
$\Lambda_\pm=\big|\mathbb{S}_{\chi_{-S}\chi_S}^+\big|^2\pm\big|\mathbb{S}_{\chi_{-S}\chi_S}^-\big|^2$.

Finally, before we write the final expression for the kinetic coefficient, let's note that for a half-integer spin impurity
(in the absence of an external magnetic field)  $\mathbb{S}_{\chi_S\chi_{-S}}^z=\mathbb{S}_{\chi_{-S}\chi_S}^z=0$
and $\mathbb{S}_{\chi_{\pm S}\chi_{\pm S}}^\pm=0$, which allows for the following identities to be used in Eq.~(\ref{Eq:kin_coeff_app}),
    \begin{align}
    \sum_{\chi\chi^\prime}
    \mathcal{W}_{\sigma\sigma^\prime}^{\chi\chi^\prime}
     \Bigg(
    \frac{\partial \mathcal{P}_\chi}{\partial x_k}\Big|_\text{eq}
     -
    \frac{\partial \mathcal{P}_{\chi^\prime}}{\partial x_k}\Big|_\text{eq}
    \Bigg)
    =
    \big(\alpha_\text{ex}^{LR}\big)^{\!2}
    \delta_{\sigma^\prime\overline{\sigma}}
    \eta_\sigma
    \Lambda_-
     \Bigg(
    \frac{\partial \mathcal{P}_{\chi_S}}{\partial x_k}\Big|_\text{eq}
     -
      \frac{\partial \mathcal{P}_{\chi_{-S}}}{\partial x_k}\Big|_\text{eq}
     \Bigg)
     ,
    \end{align}
and
    \begin{align}
    \sum_{\chi\chi^\prime}
    \big|\mathbb{S}_{\chi^\prime\chi}^+\big|^2
     \Bigg(
    \frac{\partial \mathcal{P}_\chi}{\partial x_k}\Big|_\text{eq}
     -
    \frac{\partial \mathcal{P}_{\chi^\prime}}{\partial x_k}\Big|_\text{eq}
    \Bigg)
    =
    \Lambda_-
     \Bigg(
    \frac{\partial \mathcal{P}_{\chi_S}}{\partial x_k}\Big|_\text{eq}
     -
      \frac{\partial \mathcal{P}_{\chi_{-S}}}{\partial x_k}\Big|_\text{eq}
     \Bigg)
     .
    \end{align}
Accordingly, the formulas for the kinetic coefficients take the general forms
    \begin{align}
    \hspace*{-5pt}
     \mathcal{L}_{nk}=\frac{\pi}{\hbar}
     K^2
     \Bigg[
     &
     \sum_{\chi\chi^\prime}
    \sum_{\sigma\sigma^\prime}
    (\eta_\sigma\delta_{\sigma\sigma^\prime})^{\delta_{n1}+\delta_{k1}}
    \mathcal{W}_{\sigma\sigma^\prime}^{\chi\chi^\prime}
    \phi_{\sigma\sigma^\prime}^{(\delta_{n2}+\delta_{k2},0)LR}
                            \nonumber\\
     -\ &(\delta_{n0}+\delta_{n2})
     ( \delta_{k0} + \delta_{k2})
     \frac{\Lambda_-^2}{\Lambda_+}
     \Omega
    T
    \big(\alpha_\text{ex}^{LR}\big)^{\!4}
    \sum_{\sigma\sigma^\prime}
    \eta_\sigma
    \eta_{\sigma^\prime}
     \phi_{\sigma\overline{\sigma}}^{(\delta_{n2},0)LR}
     \phi_{\sigma^\prime\overline{\sigma^\prime}}^{(\delta_{k2},0)LR}
                            \nonumber\\
     -\ &
     \Big[
     (\delta_{n0}+\delta_{n2})
     \delta_{k1}
     +
     \delta_{n1}
    (\delta_{k0} + \delta_{k2})
     \Big]
    \frac{\Lambda_-^2}{\Lambda_+}
    \Omega
    T
     \big(\alpha_\text{ex}^{LR}\big)^{\!2}
    \sum_{q\sigma}
    \eta_q
    \eta_\sigma
    \big(\alpha_\text{ex}^{qq}\big)^{\!2}
     \phi_{\sigma\overline{\sigma}}^{(\delta_{n2}+\delta_{k2},0)LR}
     \phi_{\uparrow\downarrow}^{(0,0)qq}
                    \nonumber\\
    -\ &
    \delta_{n1}
    \delta_{k1}
    \Bigg\{
    \frac{\Lambda_-^2}{\Lambda_+}
    \Omega
    T
    \sum_{qq^\prime}
    \eta_q
    \eta_{q^\prime}
     \big(\alpha_\text{ex}^{qq}\big)^{\!2}
    \big(\alpha_\text{ex}^{q^\prime q^\prime}\big)^{\!2}
     \phi_{\uparrow\downarrow}^{(0,0)qq}
     \phi_{\uparrow\downarrow}^{(0,0)q^\prime\! q^\prime}
    -
    \Lambda_+
    \sum_q
     \big(\alpha_\text{ex}^{qq}\big)^{\!2}
    \phi_{\uparrow\downarrow}^{(0,0)qq}
    \Bigg\}
     \Bigg]
     .
     \hspace*{-3pt}
    \end{align}
As previously, it can also be easily checked that the Onsager relation holds.

\twocolumngrid


%


\end{document}